\documentclass[iop]{emulateapj}

\usepackage{natbib}
\usepackage{graphicx}

\newcommand{\mjybeam}{mJy beam$^{-1}$}

\begin{document}

\shorttitle{AU Mic: Herschel and JCMT resolved imaging}
\title{The AU Mic Debris Disk: far-infrared and submillimeter resolved imaging}

\author{Brenda C.~Matthews\altaffilmark{1,2},
Grant Kennedy\altaffilmark{3},
Bruce Sibthorpe\altaffilmark{4},
Wayne Holland\altaffilmark{5,6},
Mark Booth\altaffilmark{7},
Paul Kalas\altaffilmark{8,9},
Meredith MacGregor\altaffilmark{10},
David Wilner\altaffilmark{10},
Bart Vandenbussche\altaffilmark{11},
G\"oran Olofsson\altaffilmark{12},
Joris Blommaert\altaffilmark{13,14},
Alexis Brandeker\altaffilmark{12},
W.R.F.~Dent\altaffilmark{15},
Bernard L.~de Vries\altaffilmark{12,16},
James Di Francesco\altaffilmark{1,2},
Malcolm Fridlund\altaffilmark{17,18},
James R.~Graham\altaffilmark{8},
Jane Greaves\altaffilmark{19},
Ana M.~Heras\altaffilmark{20},
Michiel Hogerheijde\altaffilmark{18},
R.\,J.~Ivison\altaffilmark{21,5},
Eric Pantin\altaffilmark{22},
and 
G\"oran L.~Pilbratt\altaffilmark{20}}

\altaffiltext{1}{National Research Council of Canada Herzberg Astronomy \& Astrophysics Programs, 5071 West Saanich Road, Victoria, BC, Canada, V9E 2E7}
\altaffiltext{2}{University of Victoria, 3800 Finnerty Road, Victoria, BC, V8P 5C2, Canada}
\altaffiltext{3}{Institute of Astronomy, University of Cambridge, Madingley  Road, Cambridge, CB3 0HA, UK}
\altaffiltext{4}{SRON Netherlands Institute for Space Research, Groningen, The Netherlands}
\altaffiltext{5}{Institute for Astronomy, University of Edinburgh, Royal Observatory, Blackford Hill, Edinburgh EH9 3HJ, UK}
\altaffiltext{6}{UK Astronomy Technology Centre, Science and Technology Facilities Council, Royal Observatory, Blackford Hill, Edinburgh, EH9 3HJ, UK}
\altaffiltext{7}{Instituto de Astrofísica, Pontificia Universidad Cat\'olica de Chile, Vicuña Mackenna 4860, 7820436 Macul, Santiago, Chile}
\altaffiltext{8}{Department of Astronomy, University of California, 601 Campbell Hall, Berkeley, CA, 94720, USA}
\altaffiltext{9}{SETI Institute, Mountain View, CA 94043, USA}
\altaffiltext{10}{Harvard-Smithsonian Center for Astrophysics, 60 Garden Street, Cambridge, MA 02138 USA}
\altaffiltext{11}{Institute of Astronomy KU Leuven, Celestijnenlaan 
200D, 3001 Leuven, Belgium}
\altaffiltext{12}{Department of Astronomy, AlbaNova University Centre, Stockholm University, SE-106 91 Stockholm, Sweden}
\altaffiltext{13}{Astronomy and Astrophysics Research Group, Department of Physics and Astrophysics, Vrije Universiteit Brussel, Pleinlaan 2, 1050 Brussels, Belgium}
\altaffiltext{14}{Flemish Institute for Technological Research (VITO), Boeretang 200, 2400 Mol, Belgium}
\altaffiltext{15}{Joint ALMA Observatory, Alonso de Cordova 3107, Vitacura 763-0355, Santiago, Chile}
\altaffiltext{16}{Stockholm University Astrobiology Centre, SE-106 91 Stockholm, Sweden}
\altaffiltext{17}{Department of Earth and Space Sciences, Chalmers University of Technology, Onsala Space Observatory, 439 92, Onsala, Sweden}
\altaffiltext{18}{Leiden Observatory, Leiden University, Postbus 9513, 2300 RA, Leiden, The Netherlands}
\altaffiltext{19}{School of Physics and Astronomy, University of St Andrews, North Haugh, St Andrews, Fife KY16 9SS, UK}
\altaffiltext{20}{ESA, Scientific Support Office, Directorate of Science and Robotic Exploration, European Space Research and Technology Centre (ESTEC/SRE-S), Keplerlaan 1, 2201 AZ Noordwijk, The Netherlands}
\altaffiltext{21}{European Southern Observatory, Karl Schwarzschild Strasse 2, Garching, Germany}
\altaffiltext{22}{Laboratoire AIM, CEA/DSM - CNRS - Université Paris Diderot, IRFU/SAp, 91191 Gif sur Yvette, France}

\begin{abstract}
We present far-infrared and submillimeter maps from the {\it Herschel
Space Observatory} and the James Clerk Maxwell Telescope of the debris
disk host star AU Microscopii. Disk emission is detected at 70,
160, 250, 350, 450, 500 and 850 \micron. The disk is resolved at 70, 160 and 450
\micron.  In addition to the planetesimal belt, we
detect thermal emission from AU Mic's halo for the first time. In
contrast to the scattered light images, no asymmetries are evident in
the disk.  The fractional luminosity of the disk is $3.9 \times
10^{-4}$ and its mm-grain dust mass is $0.01 \ M_\oplus$ ($\pm
20$\%). We create a simple spatial model that reconciles the disk SED
as a blackbody of $53 \pm 2$ K (a composite of 39 and 50 K
components) and the presence of small (non-blackbody) grains which
populate the extended halo.  The best fit model is consistent with the
``birth ring'' model explored in earlier works, i.e., an edge-on dust
belt extending from 8.8-40 AU, but with an additional halo component
with an $r^{-1.5}$ surface density profile extending to the limits of
sensitivity (140 AU).  We confirm that AU
Mic does not exert enough radiation force to blow out grains. For
stellar mass loss rates of 10-100x solar, compact (zero porosity)
grains can only be removed if they are very small; consistently with
previous work, if the porosity is 0.9, then grains approaching 0.1
\micron\ can be removed via corpuscular forces (i.e., the stellar
wind).
\end{abstract}


\section{Introduction}

Debris disks are one of the most prevalent signposts that a stellar
system succeeded in building up planetary scale bodies during the
protoplanetary disk phase. Debris disks are collisionally sustained
distributions of planetesimals, smaller rocky bodies, and dust around
main sequence (and more evolved) stars. Because dust grains can be
removed from the system through various physical processes, their
presence is direct evidence of an unseen population of larger
planetesimals, and potentially planets, in orbit around the star.  At
current sensitivity levels, debris disks are found around $\gtrsim
20$\% of nearby solar and A-type stars \citep[][Matthews et al.\ in
prep]{eiroa-et-al-2013,thureau13,matthews14_ppvi}, with enhanced
detection rates around younger stars, e.g., the Pleiades:
\cite{gorlova06} find 25\% for B and A-type stars while
\cite{sierchio10} find 32\% for solar types; the 10-20 Myr old Sco-Cen
association: \cite{chen11} and \cite{chen12} find rates from
$25-33$\%, varying with spectral type; and ensembles of A star
populations: \cite{su06} find a rate of 32\% at 24 and 70 \micron.
There is, however, a relative paucity of disks detected around M stars
\citep{matthews14_ppvi}.  Recent surveys for debris disks have
revealed very low detection rates for M star hosted debris disks
compared to earlier type stars \citep{low05,gautier07}. For example,
data from the DEBRIS survey with the {\it Herschel Space
Observatory}\footnote[23]{{\it Herschel} is an ESA space observatory
with science instruments provided by European-led Principal
Investigator consortia and with important participation from NASA.}
reveal just two disks in a sample of 89 observed M stars within 8.6 pc
\citep[][in prep]{matthews15}.  The mass sensitivity to M star disks
in existing surveys, however, has not yet matched those of earlier
spectral types \citep[see Fig.\ 2 of ][]{matthews14_ppvi}.

Several factors affect the detectability of disks around M stars.  For
example, grain removal by stellar winds is more efficient around M
stars \citep[e.g.,][]{plavchan-et-al-2005}, suggesting that small
grains (i.e., $< 1$ \micron) may be removed from these systems at a
rate higher than expected purely from radiation forces
\citep{augereau-beust-2006,matthews07}. This effect may explain why
the highest detection rate of $13^{+6}_{-8}$\% is found for a
combination of submillimeter studies since these are sensitive to
larger grains \citep{lestrade06}. \cite{forbrich08} find that for the
$30-40$ Myr old cluster NGC 2547, the detection rates of M star disks
at 24 \micron\ exceeds that of G and K stars of the same age,
suggesting that at least around very young stars, M star disks may be
just as detectable as disks around earlier type stars, consistent with
the bright disk detected around the M3IVe \citep{pecaut-mamajek-2013}
star AU Mic, the M3IVe \citep{pecaut-mamajek-2013} star TWA 7's disk
\citep{matthews07}, and the recently identified disk
\citep{kennedy2013fc} around the M4 star Fomalhaut C
\citep{mamajek13}.  Nevertheless, a disk of high fractional luminosity
has also been detected and resolved around the significantly older,
multiple planet host GJ 581 \citep{lestrade12}. Therefore, AU Mic, as
the first and youngest nearby M star to have a detected disk, retains
particular importance as a representative of its class.

AU Mic (GJ 803, HD 197481) was the only M star detected to have a
far-IR excess above its stellar photosphere (indicative of
circumstellar dust) with IRAS \citep[Faint Source
Catalogue,][]{moshir1990}. A submillimeter excess was detected with
photometric-mode observations with the SCUBA camera, and the emission
was not resolved in a relatively shallow mapping observation
\citep{liu04}. An excess was also seen with the CSO at 350 $\mu$m
\citep{chen05}. Ground-based imaging by \cite{kalas04} and subsequent
high-resolution imaging and polarimetry from the {\it Hubble Space
Telescope} by \cite{krist05} and \cite{graham07} revealed it to be the
second debris disk spatially resolved at optical wavelengths, hosting
an edge-on disk that extends to over 100 AU in radius. The
\cite{graham07} HST polarization study revealed evidence of a change
in the polarization properties at a radius of $\sim 35$ AU and a
dearth of micron-sized grains interior to 40 AU. \cite{wilner12}
imaged the disk with the SubMillimeter Array (SMA) and most recently,
imaging with the Atacama Large Millimeter/submillimeter Array (ALMA)
yielded a well-resolved disk image modeled as a narrow ``birth ring''
or ``parent belt" of planetesimals at 40 AU \citep{macgregor13}. The
ALMA data do suggest a reasonably wide dust belt extending inward to
$\sim 9$ AU though the dust surface density is strongly peaked near 40
AU and the inner edge is poorly constrained.  In recent STIS
observations,
\cite{schneider14} report an ``out-of-plane bump'' on one side of the
disk at $\sim 13$ AU, interpreting this as a dust density enhancement,
in contrast to the ALMA data that revealed no significant asymmetries.
\cite{schneider14} also observe a brightness asymmetry between
the two sides of the scattered light disk interior to 15-20 AU, a
region that had not been cleanly imaged in earlier scattered light
detections.  Finally, the ALMA data also revealed an unresolved
excess at the position of the star which MacGregor et al.\ note could
be attributable to unresolved emission from an asteroid-like warm belt
near the star. \cite{cranmer13}, however, suggest coronal thermal
heating could alternatively account for the observed excess at the
position of the star.

AU Mic attains additional importance as a member of the Beta Pictoris
Moving Group since, as such, it has a relatively well-established
young age of $23\pm3$ Myr
\citep{binks14,malo14,mamajek14}, revised upward from $12^{+8}_{-4}$ Myr
\citep{zuckerman01}.  Due to its youth and proximity \citep[$9.91 \pm 0.10$
pc,][]{vanleeuwen07}, AU Mic is a favored target for study.  As
described in detail by \cite{wilner12}, its disk also shares many
qualitative characteristics with the $\beta$ Pictoris disk, including
its edge-on geometry and extended scattered light emission
\citep{kalas04}. The age of AU Mic is the currently favored epoch for
the formation of terrestrial planets \citep{chambers14,raymond12}.
While M stars do host planetary systems, including the multiple
planets around GJ 876 \citep{rivera05}, GJ 581 \citep{udry07} and GJ
676A \citep{anglada12}. No planets have been detected around AU Mic,
although \cite{schneider14} do suggest a potential planetary origin
for the observed asymmetry in scattered light.

AU Mic is similar to other bright debris disks in that it possesses a
bright ``halo", seen in scattered light images. In debris disk
systems where the smallest grains are blown out of the system as soon
as they are created, and larger grains are unaffected, there
necessarily exists an intermediate size range where newly created
particles are placed on eccentric orbits. The specific sizes of these
particles depends on the stellar luminosity and mass loss rate, but in
general they are smaller than 10 $\mu$m. These particles have
pericenters within the parent belt (or birth ring), but apocenters
extending to the maximum allowed by the local environment. Thus, they
form a small-grain ``halo" that surrounds the parent belt. In the case
of AU Mic, the radiation force is relatively weak and the halo is
created by the radial force exerted on the dust by the stellar wind,
which is thought to be 10-100 times greater than the Solar wind
\citep{strubbe-chiang-2006,augereau-beust-2006}. A key prediction is
therefore that halos should be relatively faint at mm wavelengths
where the small grains emit inefficiently, and thus only the parent
belt is detected. This hypothesis has been confirmed by mm-wave
observations of AU Mic \citep{wilner12,macgregor13}, as well as for
other systems \citep[e.g., Vega, $\beta$ Pic, and
HR~8799;][respectively]{holland98,williams-andrews-2006,dent13}, thus
explaining how debris disks can have different apparent sizes at
different wavelengths. For both Vega and HR~8799 the halo has only
been detected in mid/far-IR emission
\citep{sibthorpe10,su09,matthews14}.

\begin{deluxetable*}{ccccc}
\tabletypesize{\scriptsize}
\tablecaption{Observations Log}
\tablewidth{0pt}
\tablehead{
\colhead{Obs. ID} & \colhead{Observing date} & \colhead{Mode} & \colhead{Duration [s]} & \colhead{PWV [mm]}}
\startdata
\multicolumn{5}{c}{Herschel Space Observatory} \\
\hline
1342196038 & 9 May 2010 & PacsPhoto 70/160 (scan 135) & 5478 & $-$\\
1342196103 & 9 May 2010 & PacsPhoto 70/160 (scan 45) & 5478 & $-$\\
1342193786 & 5 April 2010 & SpirePhoto & 2906 & $-$\\ 
\hline
\multicolumn{5}{c}{James Clerk Maxwell Telescope} \\
\hline
SCUBA-2 & 22 April 2012 & daisy scan 850/450 & 3910 & 1.06 \\
& 23 April 2012 & daisy scan 850/450 & 3947 & 0.83 - 0.92  \\
& 17 May 2012 & daisy scan 850/450 & 1960 & 0.92 \\
& 7 June 2012 & daisy scan 850/450 & 1895 & 1.4  \\
& 16 August 2012 & daisy scan 850/450 & 3787 & 0.86 \\ 
& 19 August 2012 & daisy scan 850/450 & 3790 & 0.83 
\enddata
\label{obslog}
\end{deluxetable*}

We present the first resolved far-infrared images of AU Mic's debris
disk from the {\it Herschel Space Observatory's} PACS
\citep[Photodetector Array Camera and Spectrometer,][]{poglitsch10}
instrument at 70 \micron\ and 160 \micron\ and lower resolution
submillimeter images from the SPIRE \citep[Spectral and Photometric
Imaging REceiver,][]{griffin10} camera at 250 \micron, 350 \micron\
and 500 \micron.  These data were taken during Guaranteed Time, as
part of the Disk Evolution Key Program (PI: Olofsson).  In addition,
we present 850 \micron\ and 450 \micron\ data from the SCUBA-2
Observations of Nearby Stars (SONS) Survey and a PI program, taken
with the SCUBA-2 camera \citep{holland13}. We describe our
observations and data reduction in $\S$ \ref{obs} and present the data
in $\S$ \ref{results}. In $\S$ \ref{disc}, we discuss the new
constraints these data place on models of the disk emission, namely
the temperature of the planetesimal belt and halo and the extent of
the halo in thermal emission. We summarize our conclusions in $\S$
\ref{conc}.

\section{Observations}
\label{obs}

\begin{figure}
\includegraphics[trim=0cm 0cm 0cm 12cm,clip=true,scale=0.4]{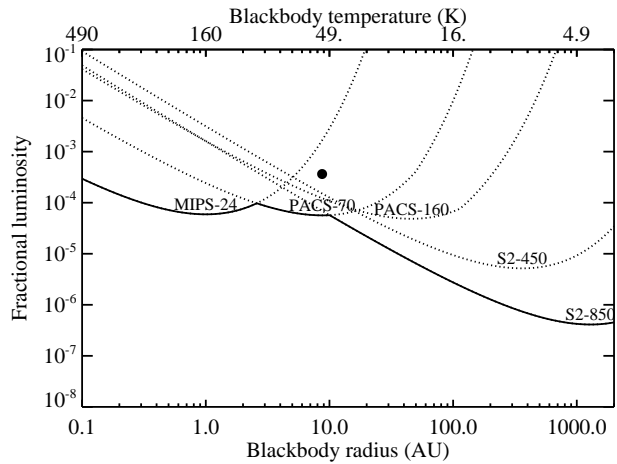}
\caption{Relative sensitivities of {\it Spitzer} MIPS (24 \micron), {\it Herschel} PACS
  and SCUBA-2 to dust emission at the level of AU Mic's disk, shown as
  a black dot. Any disk above each instrument's sensitivity curve will
  be detectable to that instrument. The best fit temperature and
  fractional luminosity of AU Mic render undetectable to MIPS at 24
  \micron\ \citep[as reported by][]{chen05}, but detectable to PACS at
  70 and 160 \micron\ as well as SCUBA-2 at 450 and 850 \micron.}
\label{sensitivity}
\end{figure}

A summary of details of the observations from {\it Herschel}
\citep{pilbratt10} and the James Clerk Maxwell Telescope (JCMT) is
shown in Table \ref{obslog}.  The relative sensitivities of {\it
Spitzer}/MIPS24, {\it Herschel}/PACS and SCUBA-2 observations to the
AU Mic disk in terms of fractional luminosity ($L_{\rm disk}/L_\star$)
against blackbody radial dust location are shown in Figure
\ref{sensitivity}.

\subsection{{\it Herschel} Observations}

\begin{figure*}
\includegraphics[trim=0cm 0cm 0cm 0cm,clip=true,scale=1.0]{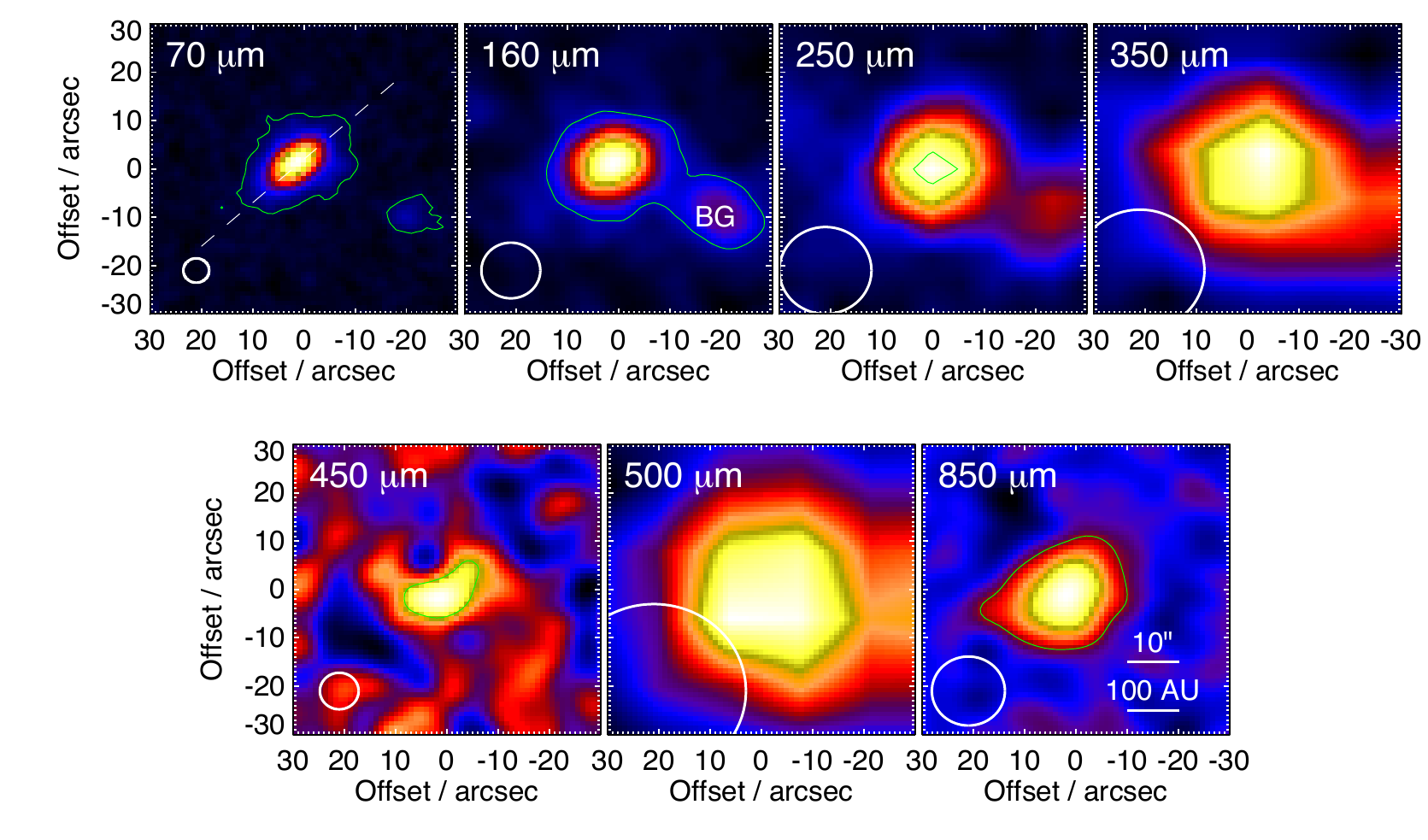}
\caption{Far-infrared and submillimeter maps of AU Mic from {\it Herschel} and
  SCUBA-2. North is up and east is to the left.  Different surface
  brightness scales are used for each map, and the pixel scales used
  are 1\arcsec\ at 70 \micron, 2\arcsec\ at 160 \micron, and 6\arcsec,
  10\arcsec\ and 14\arcsec\ at 250 \micron, 350 \micron\ and 500
  \micron, respectively.  The pixel scale of the JCMT maps is
  1\arcsec, and the maps are smoothed to a half- (850 \micron) or
  full-width (450 \micron) Gaussian. The green contours show the
  $3-\sigma$ level in each of the maps. The $1-\sigma$ rms levels are
  0.9 \mjybeam, 9.0 \mjybeam, 5.8 \mjybeam, 6.3 \mjybeam, 6.0
  \mjybeam, 6.8 \mjybeam\ and 0.9 \mjybeam\ from 70 through 850
  \micron. The rms levels for the SPIRE data are the confusion limits
  of the instrument ({\it Herschel} Observers' Manual). The background
  source, ``BG'', is well isolated from the AU Mic disk emission at 70
  \micron\ and surrounded by a $3-\sigma$ contour at 70 \micron\ and
  160 \micron\ (labeled).}
\label{images}
\end{figure*}

Large area photometric mapping observations were performed using the
PACS and SPIRE cameras at 70 \micron, 160 \micron, 250 \micron, 350
\micron\ and 500\,$\mu$m.  The large scan-map mode was used and nearly
perpendicular cross-linked scans were performed with both instruments.
The total observing time was 3.8 hours. The images at each observed
wavelength are shown in Figure \ref{images}.

The PACS observations used scan legs of 7.4\arcmin\ with a scan-leg
spacing of 38\arcsec. Medium scan rate maps containing 11 scan legs
were repeated 11 times in each scanning direction to achieve the
required map depth.  Data were obtained at both 70
\micron\ and 160 \micron\ simultaneously.

The nominal cross-linked large map observing parameters were used for
the SPIRE observations, with a smaller map of 4 arcmin by 4 arcmin
being executed.  In total, ten repeat maps were used to reach the
required depth.

The data were processed using version 13.0 of the Herschel interactive
pipeline environment \citep[HIPE,][]{ott10}. The standard pipeline
processing steps were used for both the PACS and SPIRE data. Versions
69 and 13.1 of the PACS and SPIRE calibration products were applied
respectively. The final maps have pixel scales of 1, 2, 6, 10 and
14\arcsec\ at 70, 160, 250, 350 and 500 \micron\ respectively.

As part of the data processing, the PACS time-lines were first masked
at the source location, as well as in other areas of bright emission,
and then high-pass filtered to reduce the impact of 1/f noise. The
filtered data were then converted to maps using the 'photProject'
task. Maps were likewise made from the SPIRE data using the
'naiveMapper' task. No filtering was performed on the SPIRE data.

Observations of $\alpha$ Tau and Neptune, adjusted to the correct
spacecraft position angle, were used as the model PACS and SPIRE PSFs
respectively, and used for model image convolution.

\subsection{JCMT Observations}

Observational data presented in this paper were also taken using the
SCUBA-2 camera \citep{holland13} on the JCMT. The data were obtained
both as part of the SONS JCMT Legacy survey \citep{phillips10} and a
PI program. The observations were taken with the constant speed DAISY
pattern \citep{holland13}, which maximizes exposure time and provides
uniform coverage in the central 3\arcmin\ diameter region of a
field. The total integration time was just over 5 hours, split into 10
separate $\sim 30$ minute observations. Observing conditions were
generally excellent with precipitable water vapour levels less than 1
mm, corresponding to zenith sky opacities of around 1.0 and 0.2 at 450
and 850 microns respectively (equivalent to JCMT weather ``grade 1'';
$\tau_{\rm 225 GHz}$ of $< 0.05$). The data were calibrated in flux
density against the primary calibrator Uranus and also secondary
calibrators CRL 618 and CRL 2688 from the JCMT calibrator list
\citep{dempsey13}, with estimated calibration uncertainties amounting
to 10\% at 450 \micron\ and 5\% at 850 \micron.

The SCUBA-2 data were reduced using the Dynamic Iterative Map-Maker
within the STARLINK SMURF package \citep{chapin13} called from the
ORAC-DR automated pipeline \citep{cavanagh08}.  The map maker used a
configuration file optimized for known position, compact sources. It
adopts the technique of ``zero-masking'' in which the map is
constrained to a mean value of zero (in this case outside a radius of
60\arcsec\ from the center of the field), for all but the final
interation of the map maker \citep{chapin13}.  The technique not only
helps convergence in the iterative part of the map-making process but
suppresses the large-scale ripples that can produce ringing artefacts.
The data are also high-pass filtered at 1 Hz, corresponding to a
spatial cut-off of $\sim 150$\arcsec\ for a typical DAISY scanning
speed of 155\arcsec/s. The filtering removes residual low-frequency
(large spatial scale) noise and, along with the ``zero-masking''
technique, produces flat, uniform final images largely devoid of
gradients and artefacts \citep{chapin13}.

To account for the attenuation of the signal as a result of the time
series filtering, the pipeline re-makes each map with a fake 10 Jy
Gaussian added to the raw data, but offset from the nominal map center
by 30\arcsec\ to avoid contamination with any detected source. The
amplitude of the Gaussian in the output map gives the signal
attenuation, and this correction is applied along with the flux
conversion factor derived from the calibrator observations. The final
images were made by coadding the 10 maps using inverse-variance
weighting, re-gridded with 1-arcsec pixels at both wavelengths. The
final images at both wavelengths have been smoothed with a 7\arcsec\
FWHM Gaussian to improve the signal-to-noise ratio. The FWHMs of the
primary beam are 7.9\arcsec\ and 13.0\arcsec\ at 450 \micron\ and 850
\micron, respectively.

\begin{figure}
\includegraphics[trim=2cm 7cm 2cm 7cm,clip=true,scale=0.45]{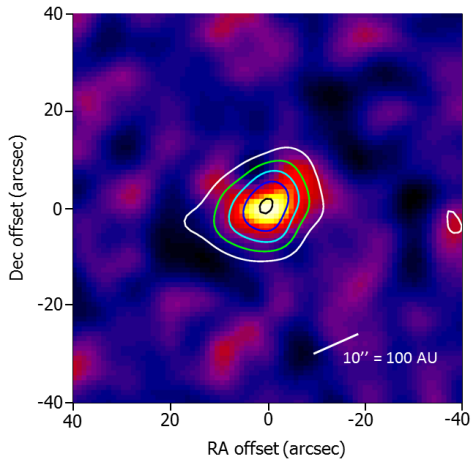}
\caption{SCUBA-2 imaging of the AU Mic disk. North is up and east is to the
  left. Composite image showing the extension of the disk along the known disk
  orientation (shown by the orientation of the scale bar) at 450 \micron\ in greyscale with an overlay of the 850 \micron\
  contours. The contour levels are RMS levels of 3-$\sigma$ (white), 6-$\sigma$ (green),
  9-$\sigma$ (light blue), 12-$\sigma$ (blue) and 15-$\sigma$ (black). }
\label{aumic-scuba2}
\end{figure}

\begin{figure}
\begin{center}
\includegraphics[trim=0cm 5cm 0cm 5cm,clip=true,scale=0.4]{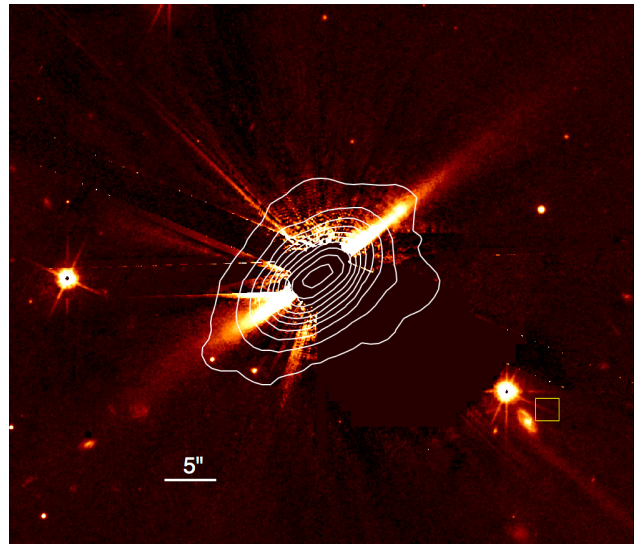}
\caption{Optical image of AU Mic from HST
  showing the orientation and extent of the scattered light
  disk. North is up and east is to the left. The 70 \micron\ emission
  from {\it Herschel} is overlaid. The disk is clearly resolved at 70
  \micron, and the disk orientations are consistent, to the limits of
  the {\it Herschel} resolution. The yellow box shows the position of
  the 70 \micron\ source detected with {\it Herschel} at the epoch of
  the HST observations. The position is close to, but not coincident
  with, the optically visible galaxy approximtely 2\arcsec\ to the
  southeast.  }
\end{center}
\label{AUMic-hst}
\end{figure}

\section{Results}
\label{results}

Figure \ref{images} shows the observed thermal emission on the sky
toward AU Mic at seven wavelengths. The maps have been cropped to show
only a 1\arcmin\ $\times$ 1\arcmin\ field centered on the star. AU Mic
is detected at all wavelengths.  The disk is resolved at 70 and 160
\micron\ with {\it Herschel}.  Two-dimensional Gaussian fits to the
JCMT data yield sizes of $16\farcs0 \times 8\farcs6$ at 450 \micron\
(at $PA \approx 135$\degr, aligned with the scattered light images)
and $16\farcs9 \times 14\farcs4$ at 850 \micron, which suggest that
the source is resolved along the major axis at 450 \micron\ (as
also indicated by the difference in the peak and integrated flux
densities in Table \ref{fluxtable}) but only marginally at 850
\micron\ (see composite Figure \ref{aumic-scuba2}).

The PACS flux densities were measured using aperture photometry,
with radii of 17\arcsec\ at 70 \micron\ and 34\arcsec\ at 160
\micron. We used aperture corrections of 0.81 and 0.85 at 70 and 160
$\mu$m, which were derived from a large set of calibration
observations processed in the same way as the data. At 160 \micron,
this aperture includes the background source, whose point source flux
was estimated in the modelling described below and has been
subtracted to estimate the AU~Mic flux density alone. The RMS noise
levels were estimated by integrating the flux within beam-sized
circular apertures ($\sim 10$ at 70 \micron), spaced by the FWHM beam,
over the central 2\arcmin\ diameter area. However, the PACS flux
densities have an uncertainty that is largely set by the instrumental
calibration, with some additional uncertainty at 160 $\mu$m to allow
for subtraction of the background source. At 70 \micron\ and 160
\micron, the absolute flux calibration accuracies are 3\% and
5\% respectively \cite[{\it Herschel} Observers' Manual;][]{balog14}.
Due to the depth of the image, the measurement errors are negligible
compared to the calibration errors, so the combined residual sum of
squares uncertainties are the same as the calibration errors.  We
attempted to measure SPIRE flux densities using PSF fits.  Our flux
extraction at SPIRE wavelengths, however, is severely impacted by our
inability to separate the flux density of the AU Mic disk from nearby
background objects, so we do not quote any fluxes. The JCMT flux
density measurements were made using 20\arcsec~radius apertures, with
the same method for uncertainty estimation. At these wavelengths,
however, the calibration uncertainty ($\sim 10$\%) is relatively
unimportant.

Table \ref{fluxtable} reports the peak flux densities and
integrated flux densities of AU Mic at each observing wavelength. As
well, we list the PSF-fit flux density of the ``BG'' background source
identified in Figure \ref{images}. These measurements are found to be
in good agreement with those derived from the image model of AU Mic
described below.

\cite{rebull08} report flux densities of $143 \pm 2$ mJy at 24
\micron, $205 \pm 8$ mJy at 70 \micron, and $168 \pm 20$ at 160
\micron. \cite{plavchan09} report updated flux densities of $155.2 \pm
3.2$ mJy and $223 \pm 26$ mJy at 24 and 70 $\mu$m. The photospheric
flux at 24 $\mu$m is $150 \pm 2$ mJy so there is no significant excess
at this wavelength. Our measured flux density at 70 \micron\ from {\it
Herschel} is consistent with the measured {\it Spitzer} flux density,
while we measure a higher flux density (but consistent within
2-$\sigma$) than {\it Spitzer} at 160 \micron.

The only significant source of emission within 1\arcmin\ of AU Mic is
located at $PA = 244.0^\circ$ and separation $\sim 25$\arcsec. 
To determine if this is a background galaxy, we checked
the HST F814W observations made with ACS/WFC in 2004 (GO-10228; PI
Kalas). Figure 4 shows the 70 \micron\ {\it Herschel}
contours overlaid on the HST image. No galaxy is detected in
these optical data that corresponds to the 70 \micron\ source to the
southwest of AU Mic (position at epoch of optical data is given by the
yellow box), although the galaxy in the optical image is quite close
($\sim 2$\arcsec, the pointing accuracy of {\it Herschel}).  The
absence of a coincident galaxy is not unexpected given the relative
sensitivity to infrared faint galaxies of {\it Herschel} compared to
optical observations.

\begin{deluxetable*}{lccccc}
\tablecaption{Measured fluxes}
\tablewidth{0pt}
\tablehead{
\colhead{Wavelength} & \colhead{Peak Flux} & \colhead{Integrated Flux} & \colhead{$F_{BG}$} & \colhead{$F_{phot}$} & \colhead{Disk Flux} \\
\colhead{[\micron]} & \colhead{[mJy beam$^{-1}$]} & \colhead{[mJy]} & \colhead{[mJy]} & \colhead{[mJy]} & \colhead{[mJy]} }
\startdata
70 & $91.6 \pm 2.7$ & $231 \pm 7$ & $11.6 \pm 0.35$ & 19.6 & $219 \pm 7$\\
160 & $176 \pm 8.8$ & $228 \pm 15$ & $47.1 \pm 2.3$ & 3.6 & $226 \pm 15$ \\
450 & $32.3 \pm 5.5$ & $49.2 \pm 8.5$ & non-detection & 0.44 & $35.4 \pm 8.5$\\
850 & $12.9 \pm 0.9$ & $14.0 \pm 1.5$ & non-detection & 0.10 & $12.5 \pm 1.5$
\enddata
\tablecomments{The columns contain (1) the observing wavelength; (2) the peak flux intensity as measured from the maps of Figure \ref{images}; the uncertainty represents the $1-\sigma$ RMS in the maps; (3) the integrated intensity measured by aperture photometry; (4) the flux of the SW background source as derived from a PSF fit, color-corrected; (5) the predicted flux density of the stellar photosphere, color-corrected; and (6) the disk flux, color-corrected. The uncertainties on the extrapolated photospheric fit are of the order of a few percent and therefore negligible compared with the measured uncertainties.}
\label{fluxtable}
\end{deluxetable*}

\begin{figure}
\includegraphics[trim=0cm 0cm 0cm 12cm,clip=true,scale=0.4]{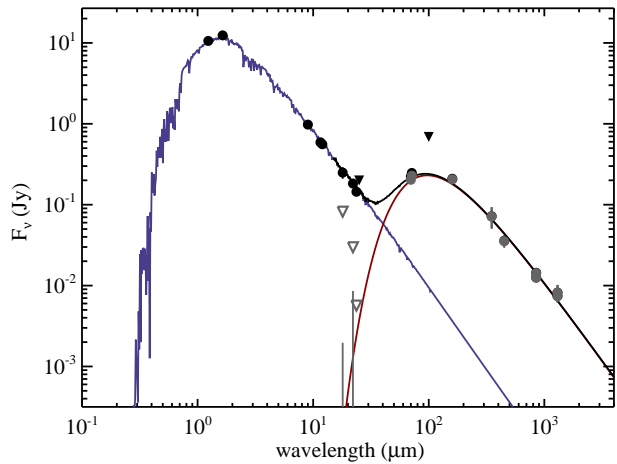}
\caption{SED of AU Mic and its disk. The fit to the stellar
  photosphere is given in blue; the disk fit is shown in red, and the
  composite stellar + disk spectrum is shown in black.  Black symbols
  are observed flux densities and grey symbols show star-subtracted
  values. Inverted triangles show upper limits (the 100 \micron\ value
  is from IRAS). In the absence of resolved imaging, a single
  component disk would be a reasonable interpretation of this SED, for
  which the excess emission is well characterized by a pure blackbody.
  }
\label{aumic-sed}
\end{figure}

The flux density distribution (which we loosely call the spectral
energy distribution, or SED) of AU Mic and its disk is shown in Figure
\ref{aumic-sed}. To derive the stellar spectrum we compiled UBV,
Hipparcos/Tycho-2, 2MASS, Spitzer, AKARI, and WISE photometry up to 12
$\mu$m. We excluded the $U-B$ color to avoid potential
issues with variability due to flaring at the shortest
wavelengths. The best fit to the BT-Settl stellar atmosphere models
\citep{allard2012} yields $T_{eff} = 3600 \pm 20$ K, a radius of 0.83
$R_\odot$, and a luminosity of 0.1 $L_\odot$. Low-res IRS Spitzer data
range from $5-14$ \micron\ and are consistent with the stellar
photosphere, but not shown on Figure \ref{aumic-sed}\footnote[24]{A
high-resolution IRS Spitzer scan was examined but was not useable.}.

We then subtracted the photospheric model from the flux densities
at longer wavelengths, and fitted a pure blackbody to this
star-subtracted photometry. The resulting model is shown in Figure
\ref{aumic-sed}, where the disk has a temperature of $53 \pm 2$ K and
a fractional luminosity of $3.5 \times 10^{-4}$.  That the disk
spectrum is well fit by a pure blackbody is surprising for two
reasons; first, most debris disks have an emission spectrum steeper
than Rayleigh-Jeans at long wavelengths
\citep[e.g.,][]{wyatt-2008,gaspar12}, and second, the disk is well
known to be extended from scattered light imaging
\citep{kalas04,liu04alone,krist05,metchev-etal-2005,fitzgerald07,graham07,schneider14}
so must comprise dust at a wide range of temperatures. The goal of the
spatial modelling in the next section is to use a simple model to
reconcile the disk extent and blackbody spectrum.

\begin{figure*}
\begin{center}
\includegraphics[scale=0.7,angle=90]{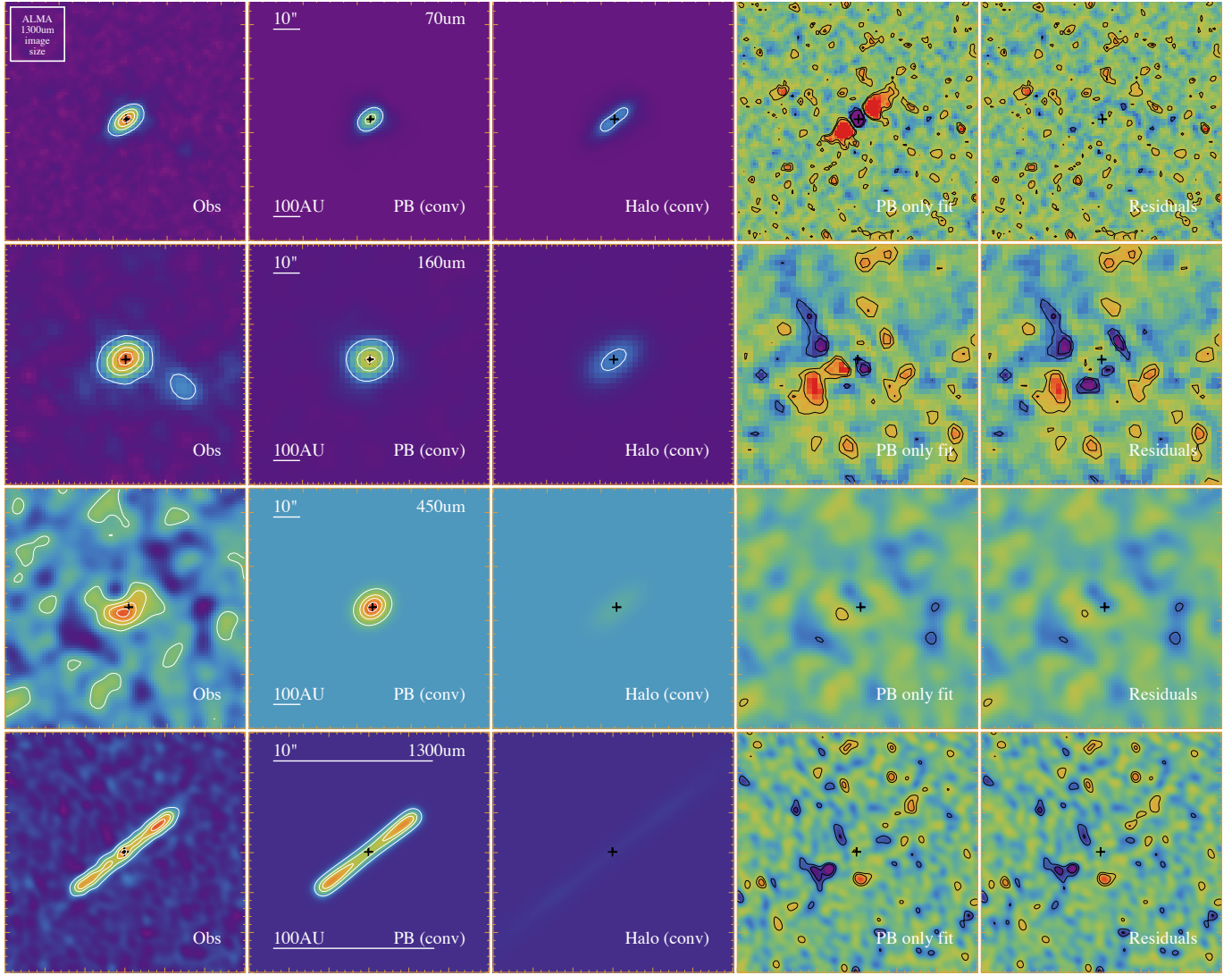}
\caption{Resolved images of the AU Mic disk at (top to bottom) 70 \micron, 160 \micron,
  450 \micron\ and 1300 \micron\ (left column, as in Figure
  \ref{images}) compared to the planetesimal belt (second column) and
  halo (third column) components. The fourth column shows residuals
  (and $\pm$2- and 3-$\sigma$ contours) when only the planetary belt model (multiplied by 1.8, 1.4, 1 and 1 for 70, 160, 450 and 1300 \micron, respectively) is removed from the data and the firth column shows the residual when the full (parent belt + halo) model is
  subtracted from the data, and includes additional background
  components, which are most visible in the 160 \micron\ image. The
  color scale for the first three panels along each row is the same to
  show the relative contribution of each component. The color scale on
  each row is different, scaled to near the peak flux in the observed
  image at that wavelength. The color scale of the residual images are
  scaled to the same $\pm \sigma$ level.  While the residuals at 450
  \micron\ are suggestively symmetric, they are not aligned with the
  known disk orientation and are barely at the 2-$\sigma$ level.  In
  addition, a point source at the stellar position was included in the
  model at 1.3 mm, consistent with the unresolved excess reported by
  \cite{macgregor13}. The residual to the south of the star at 1.3 mm
  is assumed to be an unrelated background source.  }
\end{center}
\label{fig:AUMic-model}
\end{figure*}

\begin{figure}
\includegraphics[trim=0cm 0cm 0cm 10cm,clip=true,scale=0.4]{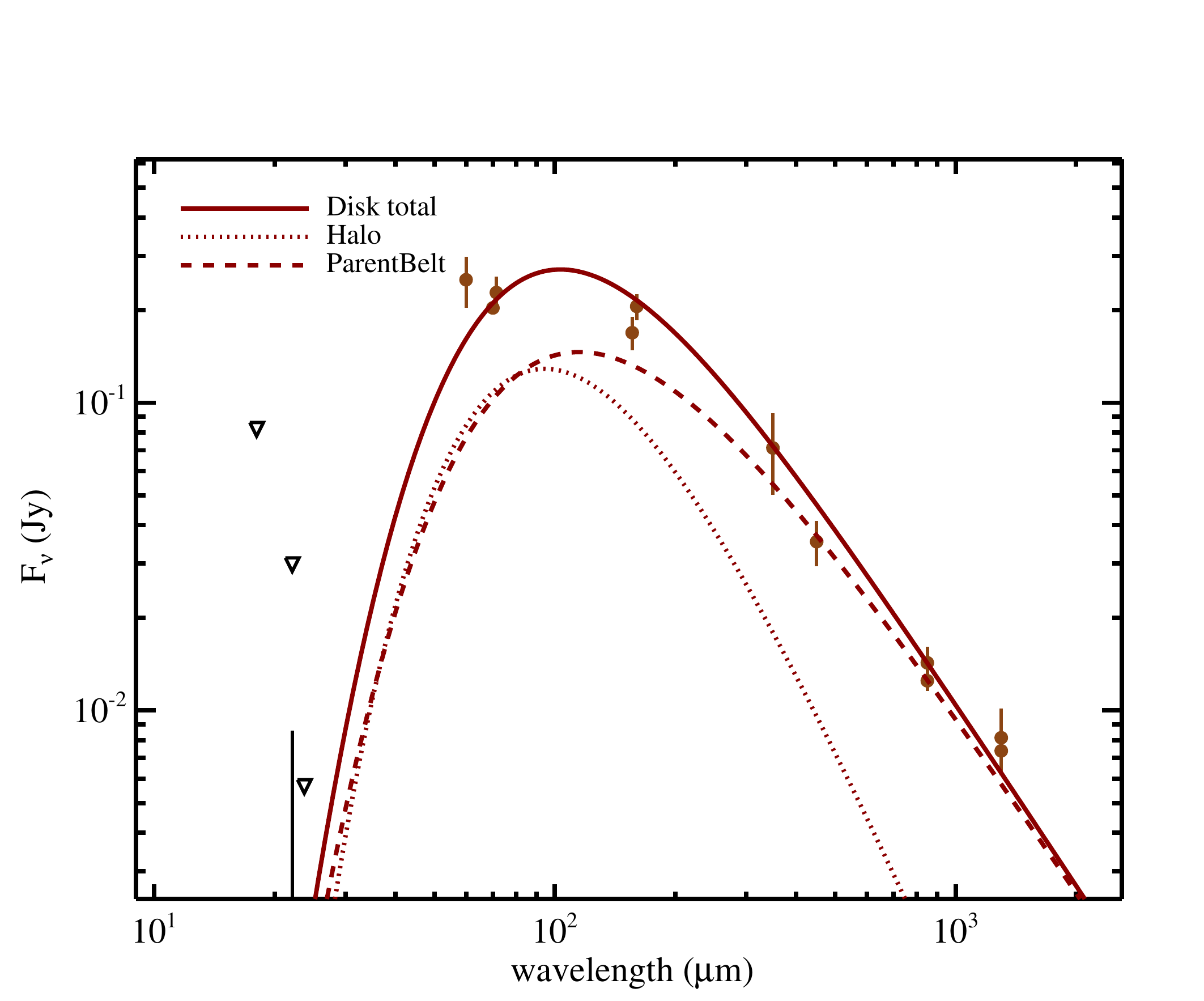}
\caption{Flux distribution of the disk around AU Mic. Subtraction of the stellar fit from
  Figure \ref{aumic-sed} reveals the distribution of the disk itself. Mid-infrared upper
  limits are shown from WISE (22 \micron), {\it Spitzer} (24 \micron) and {\it IRAS} (25
  \micron). Far-infrared detections at 60 \micron\ ({\it IRAS}) and 70 \micron\ and 160
  \micron\ ({\it Spitzer} and {\it Herschel}) are shown. Submillimeter detections are
  from the CSO (350 \micron), JCMT (450 and 850 \micron) and ALMA (1.3 mm). The model
  components of the planetesimal belt (``parent belt'') and the halo are shown. }
\label{aumic-sed-disk}
\end{figure}

\subsection{Modeling}
\label{basicmodelsec}

We use a simple model to interpret the \emph{Herschel}, SCUBA-2,
and ALMA observations. Preliminary tests find that the \emph{Herschel}
70 $\mu$m image is significantly more extended than the parent belt
imaged with ALMA at 1.3 mm, so the basic requirement of the model is
that it reconciles the different extent of these images and produces a
blackbody-like spectrum. As is already known, the solution is that the
greater radial extent at 70 $\mu$m originates in the halo of small
grains also seen in scattered light, and that these are not seen with
ALMA because these grains emit inefficiently at wavelengths
significantly larger than their physical size. Thus, our model
comprises two components; the first is the parent bodies for which we
use the ALMA modelling results of \citet{macgregor13}, and the second
is a halo whose basic properties are to be informed by previous theory
work and determined from the modelling.

With only six photometric points in the spectrum, and a only few
beams of resolution in the new \emph{Herschel} and JCMT images, our
modelling approach is physically motivated but simple. It has been
used previously to model many \emph{Herschel}-resolved debris disks
\citep[e.g.,][]{kennedy12binary,kennedy12_99her,matthews14}. For each
disk component a single azimuthally symmetric 3D dust distribution is
used, which is simply a small scale height disk with a power-law
surface density dependence between the inner and outer radii. At each
radial location, the disk emission is assumed to arise from a modified
blackbody, with a power-law radial temperature dependence. This
approach is therefore largely empirical, and the derived blackbody
parameters can be subsequently compared with more detailed dust models
to draw conclusions about the dust properties, in particular their
typical size.  The limited resolution also precludes derivation of
comparative radial profiles, as was done for HR 8799's halo
\citep{matthews14}.

Practically, we generate high resolution images at each observed
wavelength, as viewed from a specific direction to produce the desired
disk geometry. We can include an arbitrary number of disk components,
as well as point sources to model unresolved sources such as warm
inner disk regions or background sources. Each image is then convolved
with the appropriate point spread function (PSF), which may be an
observation of a bright calibration source (i.e., PACS, SPIRE), or
simply a Gaussian with specified width(s) and position angle (i.e.,
JCMT, ALMA), and then resampled to the resolution of the
observation. We model the ALMA image and not the visibilities, which
is acceptable given the good $uv$ coverage and dynamic range of the
ALMA data. The models are fit to the data using a combination of
by-eye variation of parameters and least-squares minimization. Since
we do not search all parameter space, our best-fit models are not
necessarily unique, but must be considered a reasonable
interpretation.

Our model is guided by previous modelling work.  
The parent belt is modeled using the structure derived by
\citet{macgregor13} based on ALMA data, here using a surface density
distribution between 8.8 AU and 40 AU with a radial power law
dependence of $\Sigma \propto r^{2.8}$. The inner edge is not tightly
constrained. MacGregor et al.\ find that the inner edge of the parent
belt could be as far out as 21 AU, suggesting a much narrower disk
width, though we note that \cite{schneider14} find evidence of a
stello-symmetric warp in the disk with an outer edge of 15 AU, which
supports the idea that some component of the disk has an inner edge
closer than 21 AU. Note that we are calling this component the parent
belt for convenience, and that the extent inward of 40 AU may be due
to stellar wind drag from a true source region that is much narrower,
but likely still concentrated at $\sim 40$ AU. The emission properties
of grains in this belt are assumed to be blackbody-like, and follow
$T_{\rm PB} = T_{\rm 0,PB} /\sqrt{r}$, meaning that the surface
brightness profile matches the power-law found by
\citet{macgregor13}. Because radiation and wind forces are relatively
weak for AU Mic, small dust can reside in the planetesimal belt and
$T_{\rm 0,PB}$ can be higher than expected for blackbodies, so we
leave it as a free parameter. We also add the unresolved point source
component at the stellar position seen in the ALMA data
\citep{macgregor13}, though this source is too faint to be detected at
any other wavelength (see $\S$ \ref{asteroid}).

To reproduce the {\it Herschel} 70 \micron\ image we include the
halo component, with a surface density profile fixed to $\Sigma
\propto r^{-1.5}$ \citep[e.g.,][]{strubbe-chiang-2006}, which extends
from 40 AU to 140 AU. (The outer edge is poorly constrained by the
data and fixed at 140 AU.) The surface density of this component is
forced to join smoothly to the planetesimal belt, and the grain
properties are allowed to vary via their temperatures and emission
spectra. Their spectra are modified blackbodies (i.e., a Planck
function multiplied by $(\lambda_0/\lambda)^{\beta_\lambda}$ for
$\lambda>\lambda_0$), but we reduce the number of parameters by
enforcing a temperature law suited for small grains, $T_{\rm halo} =
T_{\rm 0,halo} r^{-1/3}$. We vary $\lambda_0$,
${\beta_\lambda}$, and $T_{\rm 0,halo}$ as free parameters. As noted
above we expect $\lambda_0 \lesssim 100$ $\mu$m and
${\beta_\lambda}>0$, so that the halo is detected at 70 $\mu$m, but
not at 1.3 mm with ALMA. These values are consistent with measured
values of $\lambda_0$ and $\beta$ for other disks
\citep[i.e.,][]{booth13,matthews14}.

Finally, we include the point source visible to the west of
AU~Mic in Figure \ref{images}. This source is included in part to
ensure it does not bias the model at 160 \micron, but is also used to
derive a flux density to subtract from the 160 \micron\ aperture flux
derived above. We also include the central point source seen in the
1.3 mm ALMA image. With this simple model we have four main
parameters; one for the parent belt temperature and three for the halo
component temperature and spectrum. There are in addition eight RA/Dec
offsets for the \emph{Herschel}, JCMT, and ALMA images, but these are
relatively well constrained. The aim of the model fitting is therefore
essentially to find the relative weight of the two components at each
wavelength, thereby empirically deriving a coarse spectrum for each in
a way that is independent of assumptions about grain properties. We
found that the relatively poor resolution (excepting the ALMA data)
means that the parameters are poorly constrained due to degeneracies,
in particular for the halo component. However, all solutions we found
are sufficiently similar that the ultimate interpretation is the same.

Figure \ref{fig:AUMic-model} shows the original images, the parent
belt and halo model components, and the model-subtracted residuals
(for just the planetesimal belt and the full model) at wavelengths of
70 \micron, 160 \micron, 450 \micron\ and 1300 \micron. First, to make
it clear that the halo is detected, the fourth column shows the data
with only the parent belt subtracted. In this column the brightness of
the parent belt is 1.8 and 1.4 times brighter than in the final model
at 70 and 160 $\mu$m. The 70 $\mu$m residuals clearly show that the
disk is more extended than the parent belt, and that a component with
greater extent (i.e., the halo) is required. The fifth (rightmost)
column shows the residuals with the addition of this component, and
that our parent belt + halo model reproduces the data well. Thus, the
halo contributes significant emission at 70 $\mu$m, but little at
longer wavelengths.

At all wavelengths, few residuals above the 3-$\sigma$ level are
seen. The negative residuals at 160 $\mu$m appear to lie along the
scan directions (at roughly $\pm45^\circ$ relative to N) suggesting
that they are artefacts of the imaging. The positive residual to the
SE lies beyond the disk extent and is not seen on the opposite side of
the star, so should not be considered as a deficiency of the
model. The ALMA residual map has a bright point source offset from the
stellar position. This is a background source identified by
\cite{macgregor13} as well.  It is located $0\farcs3$ west and
$1\farcs8$ south of the star and is detected with a flux density of
0.15 mJy (a 5-$\sigma$ detection).

The main derived model parameters are $T_{\rm 0,PB}=245$ K,
$T_{\rm 0,halo}=161$ K, $\lambda_0=12$ $\mu$m, and
${\beta_\lambda}=1$. Because the structure of the parent belt is
fixed, $T_{\rm 0,PB}$ is moderately well constrained by the
contribution required at 70 $\mu$m. The halo properties, which are our
primary interest, are less well constrained because is it only
strongly detected at 70 $\mu$m. However, the conclusion that
$\lambda_0$ is shorter than $\sim$70 $\mu$m, and that ${\beta_\lambda}
\gtrsim 1$ is robust because the halo is required by the images to
have little contribution at 160 $\mu$m and beyond. Given the
conclusions of previous studies that the halo is populated by small
dust on eccentric orbits these parameters are consistent with our
expectations. The specific value of $\lambda_0=12$ \micron\ suggests
$\sim$1 \micron\ sized grains dominate the emission, but a poor
constraint on this parameter means that the grains could be larger or
smaller, though by no more than an order of magnitude. We return to
the likely size of blowout grains when considering small-grain
dynamics in $\S$ \ref{ss:dyn}.

Figure \ref{aumic-sed-disk} shows the star-subtracted SED and the disk
model, and here the relative contributions of the two components as a
function of wavelength can again be seen.  We note that the ALMA
fluxes (1300 \micron) lie somewhat above the fit, but are consistent
with the model at the $2-\sigma$ level. We discuss the spectral index
of the disk in $\S$ \ref{specindex} below. As noted above, at 70
$\mu$m the emission from the planetesimal belt and halo components is
similar, but at longer wavelengths the planetesimal belt
dominates. The planetesimal belt is dominated by emission at 40 AU,
where the grains have a temperature of 39 K. The temperature of the
halo component at that separation from the star is slightly higher (50
K), but poorly constrained due to the lack of resolution and
degeneracy with $\lambda_0$ and ${\beta_\lambda}$.  The fact that
these two components yield a composite spectrum that is a simple
blackbody is a reminder of the power of resolved imaging, since much
information can be hidden within a single component SED.

\subsection{Central Asteroid Belt?}
\label{asteroid}

\cite{macgregor13} found evidence of an excess above the photosphere 
at the position of the star in their high resolution ALMA imaging, and
we include this component in the ALMA model image. The residuals from
fitting our dust model to the {\it Herschel} and JCMT images in Figure
6 show no sign of any other unresolved excess above the stellar
photosphere, so the best we can do is place an upper limit on the
emission from such a centralized component of the point-source
sensitivity at a given wavelength \citep[i.e.,
$3-\sigma$,][]{macgregor13}. It is also possible, given the
considerably lower resolution of our observations, that such emission,
if present, has been incorporated into the belt emission and removed,
which adds uncertainty to the flux density of the belt.

From Fig.\ 4 of \cite{cranmer13}, the coronal model of AU Mic predicts
a flux of a few $\mu$Jy around 100 \micron\ with a relatively flat
distribution shortward of 1300 \micron\ where the ALMA detection was
made. Our point source sensitivity is on the order of 1 mJy at 70
\micron\ and $\sim 10$ mJy at 160 \micron, meaning that, even in the 
absence of the disk, we would not
be able to detect a coronal thermal heating contribution to the
total flux in the {\it Herschel} data.

\subsection{Mass of the Disk}

A direct estimate of the mass of dust in AU Mic's debris disk can be
made from its submillimeter flux densities. Debris disks are optically
thin at these wavelengths, and so the mass of the disk is directly
proportional to the emission, following the relation:

\begin{equation} M_d = \frac{F_\nu \ d^2}{\kappa_\nu \ B_\nu(T_d)}
\label{eq-1}
\end{equation}

In the Rayleigh-Jeans limit, equation \ref{eq-1} reduces to

\begin{equation} M_d [M_\oplus] = 5.8 \times 10^{-10} \frac{F_\nu [mJy] \ (d [pc])^2 \ (\lambda [\micron])^2}{\kappa_\nu [cm^2 g^{-1}] \ T_d [K]}
\label{eq-2}
\end{equation}

While the AU Mic SED supports a single temperature fit of 53 K, our
image modeling requires a separate planetesimal belt of 39 K and a
warmer halo, which contributes negligible flux density to the total
emission at 850 \micron\ (see Fig.\ \ref{aumic-sed-disk}).  This
temperature is comparable to the 40 K adopted by
\cite{liu04} in the absence of an SED temperature fit. The color
corrected flux density measured with SCUBA-2 at 850 \micron\ (12.5
mJy) is within the $1-\sigma$ limit of the value measured by
\cite{liu04} with SCUBA ($14.4 \pm 1.8$ mJy), although the SCUBA
measurement relied on single bolometer ``photometry'' mode, rather
than a mapping technique, which could have missed some emission.

Taking 12.5 mJy as the 850 \micron\ flux density of the disk yields a
slightly lower 1-mm grain dust mass of 0.01 $M_\oplus$ relative to the
\cite{liu04} measurement, for the same dust opacity of 1.7 cm$^2$
g$^{-1}$.  The dominant sources of uncertainty in the mass are the
flux density and the temperature of the planetesimal component of the
disk, amounting to a 20\% uncertainty in the disk mass.

\section{Discussion}
\label{disc}

We now briefly consider the dust properties in more detail, using
realistic grain models to compare the probable dust temperatures and
sizes with those derived from the modelling.

\subsection{Temperature of AU Mic Dust Grains}

\begin{figure}
\includegraphics[trim=0cm 0cm 0cm 12cm,clip=true,scale=0.4]{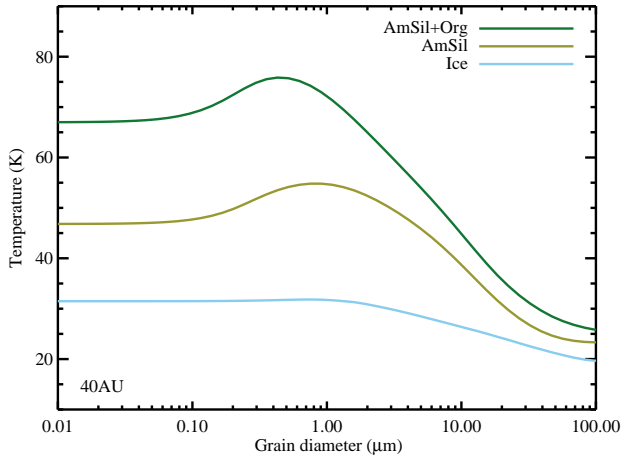}
\caption{Temperature of grains as a function of diameter for three dust grain compositions. For the planetesimal belt temperature of 39 K, pure icy grains are ruled out for all considered grain sizes for compact, spherical grains, although other compositional mixes including ices could potentially satisfy the modeled temperature. See the text for an explanation of the grain compositions. }
\label{fig:AUMic-temperature}
\end{figure}

The SED fit to the disk around AU Mic requires only a single temperature component. The
emission is well fit ($\chi^2$ is minimized) from mid-IR through submillimeter
wavelengths by a pure blackbody with a temperature of $53 \pm 2$ K.  Figure
\ref{aumic-sed-disk} shows, however, that the halo and planetesimal belt components could
still be segregated in temperature, with the parent bodies of the belt colder ($\sim 39$
K) than the halo \citep[see also ][]{fitzgerald07}.

Figure \ref{fig:AUMic-temperature} shows the distribution of
temperature with grain size for three grain compositions at the
location of the planetesimal belt. The temperature distributions were
calculated as in \cite{augereau-etal-1999} and \cite{Wyatt02}. We used
three different compositions: i) a mix of 1 part amorphous
astronomical silicates to 2 parts organics, ii) pure astronomical
silicates, and iii) pure ice \citep{li-greenberg-1997}. The models
shown in Figure \ref{fig:AUMic-temperature} are not porous. Porous
grains do not have the peak in temperature near 1 \micron, instead
decreasing steadily to $20-25$ K from maximum temperatures similar to
the non-porous case. It is clear that pure water ice grains are ruled
out at 39 K for all grain sizes, although we note that other
compositional mixes could be modeled to satisfy the temperature
distribution we have derived.

\begin{figure*}
\includegraphics[trim=0cm 0cm 0cm 12cm,clip=true,scale=0.4]{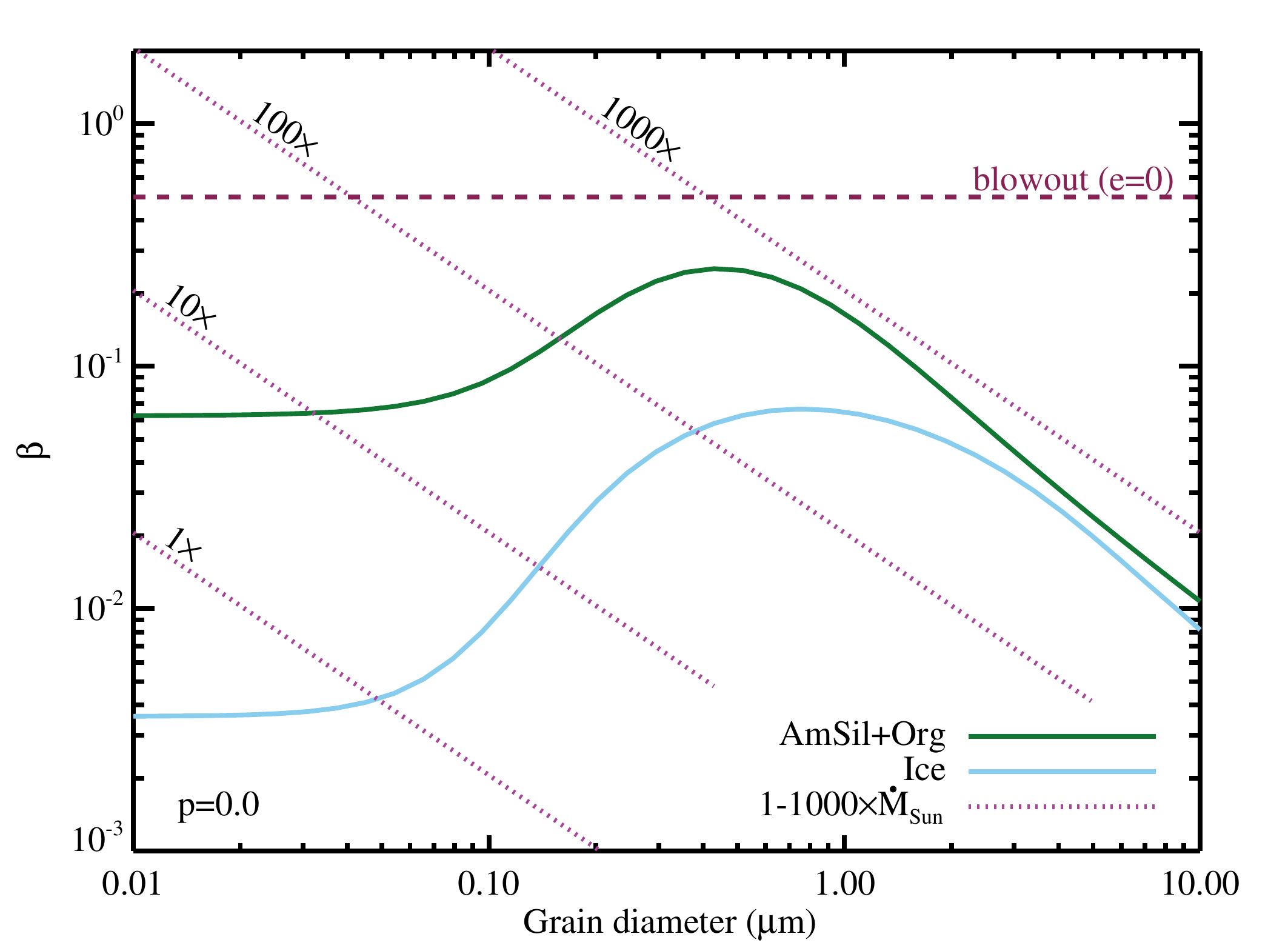}
\includegraphics[trim=0cm 0cm 0cm 12cm,clip=true,scale=0.4]{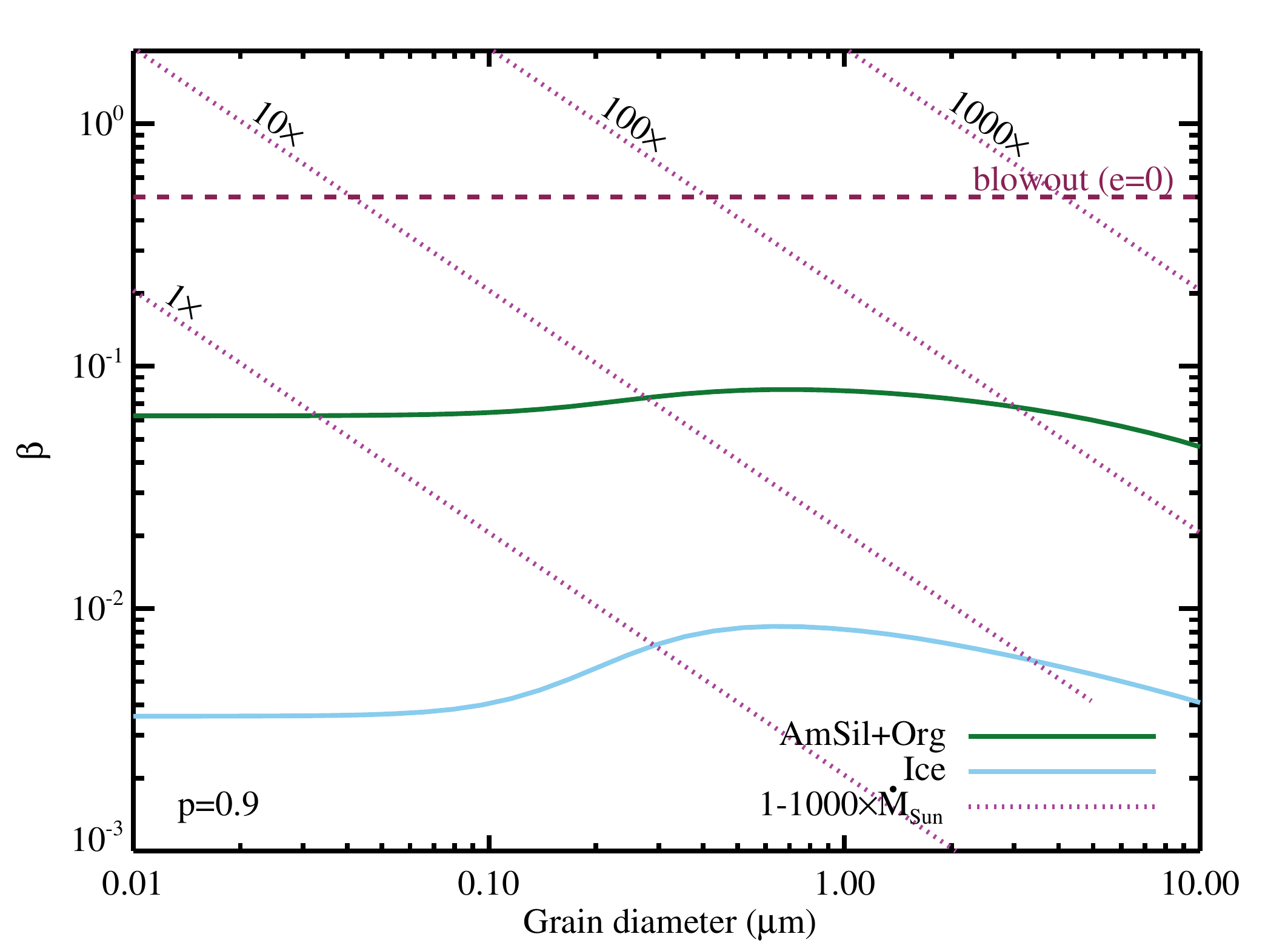}
\caption{Effects of Porosity. The ratio of radiative to gravitational forces ($\beta$) as
  a function of grain composition is shown. For zero porosity grains, there is no grain
  size that reaches the blowout condition ($\beta = 0.5$) around AU Mic, although
  non-zero porosity can either decrease or enhance the $\beta$ ratio, depending on grain
  size.  For the higher porosity grains, the $\beta$ values ultimately converge to be
  similar, following an inverse relation to grain size. Stellar wind effects on
  $\beta_{\rm SW}$ for 1, 10, $10^2$ and $10^3 \times$ $\dot{M}_\odot$ are shown as
  dotted lines. Empirical evidence for other stars suggests that AU Mic may have a
  stellar wind level as high as, but not much higher than, $10 \times \dot{M}_\odot$
  \citep{strubbe-chiang-2006}.}
\label{fig:beta}
\end{figure*}

\subsection{Small-grain dynamics}\label{ss:dyn}

Orbiting dust is subject to forces from both radiation and stellar
winds. Around early-type stars radiation dominates, while stellar
winds are thought to be more important for late-type stars such as AU
Mic. These forces are commonly split into radial and tangential
components. The radial component is called "radiation pressure" or the
"radiation force", and effectively reduces the stellar mass seen by
the particle, with an analogous effect for the stellar wind. The
tangential component is called Poynting-Robertson drag, and causes
dust grains to spiral into the star, again with an analogous (and much
stronger) effect for stellar wind. See \citet{burns79} for a detailed
review.

The small grain halo seen previously in scattered light and confirmed
here in the far-IR is similar in morphology to those seen around other
debris disks (e.g., $\beta$ Pic: \citet{pantin97,augereau01}; HR 8799:
\citet{su09,matthews14}). In other systems, the halo is attributed to
the effect of the radial component of the radiation force from the
star, which increases the eccentricity of smallest bound grains. The
key measure of this effect is the parameter $\beta$, the ratio of the
radiation force to the gravitational attraction of the star (distinct
from the mm-wave spectral slope ${\beta_\lambda}$ used above). Grains
with $\beta > 1$ are unbound, and grains liberated from a parent body
on a circular orbit are unbound if $\beta > 0.5$.  Figure
\ref{fig:beta} shows $\beta$ as a function of grain diameter, calculated for
amorphous silicates + organics, and ice as described above. The solid
curves show $\beta$ for the effect of radiation force on solid grains
(left panel) and porous grains (right panel). It is clear that a range
of $\beta$ is possible due to the poorly constrained grain
compositions. Also, the assumption of \cite{strubbe-chiang-2006} that
the radiation force blowout grain size exists (i.e., that $\beta>0.5$
for any size) is not well founded, particularly if the grains are
porous. Essentially, the low luminosity of an M type star does not
produce a strong enough radiation force to significantly affect the
small grains. Even taking into account the flaring of the star is not
enough to remove grains via the radiation force
\citep{augereau-beust-2006}.

Unlike early-type stars, an M star can have a relatively strong
stellar wind \citep{plavchan-et-al-2005}. The force from this particle
wind causes a similar effect as the radiation force and is able to
reproduce the scattered light halo around AU Mic
\citep{augereau-beust-2006}. The dotted lines in Figure \ref{fig:beta}
demonstrate the effective $\beta_{\rm SW}$ parameter for the stellar
wind force \citep[using the prescription of ][]{strubbe-chiang-2006}.
The left panel shows that for compact grains the wind blowout size is
very small, $<$0.1 $\mu$m for the favored AU Mic mass loss rate of
$10\times \dot{M}_\odot$
\citep[][based on EUV flaring]{cully94}. Such small grains are
extremely poor scatterers of light, even at optical wavelengths, so
they would appear very faint. \cite{augereau-beust-2006} get around
this problem by including flares, which increase the time-averaged
mass loss rate to $300\times \dot{M}_\odot$. Such a high mass loss
rate would however make the system transport-dominated rather than
collision-dominated, i.e., stellar wind drag would also become
important, filling the region interior to 30 AU with small dust grains
in a manner contrary to the scattered light observations
\citep{strubbe-chiang-2006}. \cite{lim96} also point out that radio
flares would be obscured if the stellar winds of M stars were many
orders of magnitude more massive than the Sun's.

The right panel of Figure \ref{fig:beta} shows that porosity can also
play an important role. By increasing the grain porosity, the effect
of wind force increases, in turn increasing the blowout grain
size. For $p=0.9$ the blowout size is an order of magnitude larger for
the same mass loss rate, and the scattering efficiency of the grains
greater, thus providing the likely solution. For a stellar wind rate
of $\sim$100 $\dot{M}_\odot$ the size of halo grains is roughly 0.1 to
1 $\mu$m. These grains have emission properties consistent with those
derived from the image modelling above. This conclusion of high
porosity compares well with \cite{graham07}, \cite{fitzgerald07} and
\cite{shen09} who all find that porosity is necessary to explain the 
scattering and polarization properties of the grains, although
\cite{shen09} find that a composition of random aggregates require a
porosity of just $\sim 0.6$ compared to the 0.9-0.94 required by
\cite{graham07} for Mie theory applied to spherical grains or
aggregates.

The \cite{strubbe-chiang-2006} model of the disk as a
collision-dominated, narrow birth ring from which the smallest grains
are blown out does a good job of explaining the halo and lack of small
grains interior to 30~AU. The recent ALMA results of
\cite{macgregor13}, however, show that there are indeed larger grains
interior to the presumed birth ring. Given the 1300 \micron\ observing
wavelength, those observations should be dominated by grains with $a
\sim \lambda/2\pi$, or 200 \micron.  Drag forces, which can act on larger
grains over long timescales, are therefore a likely cause of this
interior emission. \cite{augereau-beust-2006} concluded that
Poynting-Robertson drag is negligible for AU Mic and that stellar wind
drag may evacuate the inner regions \citep[see
also][]{strubbe-chiang-2006}. However, more recent work suggests that
stellar wind drag can fill the interior regions to some degree,
without seriously violating the scattered light constraints
\citep{schueppler2015}. Alternatively, the radially increasing surface
density of this disk is reminiscent of a ``self-stirred'' disk that
collisionally grinds down from the inside out
\citep{kenyon-bromley-2002,kennedy-wyatt-2010}. \cite{schueppler2015} also find that the self-stirred model is plausible for AU Mic's disk.

\subsection{Spectral Index at millimeter wavelengths}
\label{specindex}

The data compiled in this work provide a long lever arm to measure the
spectral index, $\alpha$ where $F_\nu \propto \lambda^{-\alpha}$ is
the emission from submillimeter through millimeter wavelengths.  As
noted above, the 1300 \micron\ fluxes from the SMA and ALMA both lie
above the nominal ``best fit'' line of the models produced by fits to
the SED and the images.  Comparisons of the 350/1300 spectral index
and the 450/1300 spectral index yields $\alpha$ values ranging from
$\sim 1.5$ to close to 2.0, which was the inferred spectral index of
350 vs. 1300 (SMA) by \cite{wilner12}. It is clear from the SED that
these wavelengths are in the Rayleigh-Jeans part of the spectrum,
which wasn't definite based on the data available to \cite{wilner12}
and \cite{gaspar12}.

A spectral index of $\sim 2$ is lower than most other debris
disks. Based on the compiled data of \cite{gaspar12}, the three A
stars in that study ($\beta$ Pictoris, Vega and Fomalhaut) have the
highest value of $\alpha$, which could suggest that the size
distribution is different around A stars, being more consistent with a
collisional cascade.  \cite{gaspar12} show that for a collisional
quasi steady-state at a single temperature, the Rayleigh-Jeans tail of
the SED should yield a spectral index of 2.65, for a self-similar
collisional differential size distribution index of 3.65.  The
shallower slopes measured by \cite{wilner12} and a comparison of the
450 \micron\ flux density from SCUBA-2 of Table \ref{fluxtable} (which
actually seems to lie somewhat below the SED fit) and the elevated
flux densities measured by the SMA and ALMA at 1300 \micron\ suggest a
size distribution index of 3.0, i.e., that the size distribution is
not the product of a collisional cascade.  The shallow slope may
however arise from the way that the parent belt and halo components
add to yield the total spectrum \citep[Fig.~7, ][]{fitzgerald07}. Such
a shallow distribution constraint has also been noted for two IR
bright disks (HD 377 and HD 104680) not detected with the VLA at 9 mm
\citep{greaves12}.  In AU Mic, it appears the grains may be very large ($>> 1$
mm). Longer wavelength observations of the disk beyond 1300 \micron\
would be very beneficial to further constrain the size distribution.

The spectral slope is also dependent on
composition. \cite{schueppler2015} present more detailed modeling of
the AU Mic disk, including investigation of the impact of
compositional dependence on the derived spectral slope.  They find a
best fit for a combination of silicate, carbon and vacuum in equal
measure, with little change in the spectral slope fit with small
additions of ice or variations of these materials. Their test also
provides clear evidence for porous grains. As in our analysis, their
SED fits also underestimate the 1300 \micron\ flux densities from SMA
and ALMA, though as in our case, this result should be interpreted
with caution since the exploration of the parameter space is limited.

\section{Conclusions}
\label{conc}

AU Mic's halo has been known since the earliest scattered light images
of the disk \citep{kalas04}, which revealed a radial extent of 210 AU.
ALMA imaging has confirmed a planetesimal belt at $\sim 40$ AU, as
predicted by \cite{strubbe-chiang-2006} and \cite{augereau-beust-2006}
and consistent with a break in the surface brightness profiles
observed in scattered light
\citep{metchev-etal-2005,fitzgerald07,graham07}. Using {\it Herschel}
and SCUBA-2 at the JCMT, we have detected and modeled the thermal
emission from both the planetesimal belt and the halo at wavelengths
of 70 \micron, 160 \micron, 450 \micron\ and 1300 \micron, the latter
being ALMA data as presented by
\cite{macgregor13}.

We present a simple spatial model that utilizes the existing model of
the planetesimal belt from ALMA imaging to reconcile the ostensibly
single-temperature blackbody SED of AU Mic's disk with the presence of
emission from the extended halo.  The best fit model is consistent
with the birth ring model explored in \cite{wilner12} and
\cite{macgregor13}, a planetesimal belt extending from 8.8-40 AU, but
with the addition of a shallow surface density profile halo
dominated by grains roughly 1 \micron\ in size.

We observe no asymmetries in the disk images, and the residual images
all show that there is negligible emission unaccounted for by a smooth
disk.

We confirm that AU Mic does not exert enough radiation force to blow
out grains. We also find that for the inferred stellar mass loss rate
of 10 times solar, compact (porosity = 0) grains can only be removed
if they are very small; if the porosity reaches 0.9 or higher, then
grains approaching 0.1 \micron\ can be removed. This result suggests
that a higher mass loss is favoured to place larger $\sim$1 $\mu$m
grains in the halo, and a high degree of porosity in the grains of AU
Mic, consistent with previous work on the scattering and polarization
properties of the disk at optical wavelengths.

The spectral index of the planetesimal belt of AU Mic may be more
shallow than our modeling suggests, if the 450 \micron\ diminished
flux density and 1300 \micron\ elevated flux densities are real. The
spectral index established from 350 \micron\ and 850 \micron\ data was
already shallow at a value of $\sim 2$, but may be as low as 1.5,
suggesting that the disk may have a grain size distribution
inconsistent with that expected of a quasi steady-state collisional
cascade.

\acknowledgements

We acknowledge the efforts of our referee, whose constructive comments
improved and clarified the manuscript.  BCM is grateful to Christine
Chen for checking the quality of high-res IRS data from {\it Spitzer}.
GMK acknowledges support from the European Union through ERC grant
number 279973.  MB acknowledges support from a FONDECYT Postdoctral
Fellowship, project no. 3140479. PK thanks support from NASA
NMO711043, NSF AST-0909188, and GO-10228 from STScI under NASA
contract NAS5-26555.  EP is partly supported by a AAS Chretien
grant. PACS has been developed by a consortium of institutes led by
MPE (Germany) and including UVIE (Austria); KU Leuven, CSL, IMEC
(Belgium); CEA, LAM (France); MPIA (Germany); INAF-IFSI/OAA/OAP/OAT,
LENS, SISSA (Italy); IAC (Spain). This development has been supported
by the funding agencies BMVIT (Austria), ESA-PRODEX (Belgium),
CEA/CNES (France), DLR (Germany), ASI/INAF (Italy), and CICYT/MCYT
(Spain).  SPIRE has been developed by a consortium of institutes led
by Cardiff University (UK) and including Univ. Lethbridge (Canada);
NAOC (China); CEA, LAM (France); IFSI, Univ. Padua (Italy); IAC
(Spain); Stockholm Observatory (Sweden); Imperial College London, RAL,
UCL-MSSL, UKATC, Univ. Sussex (UK); and Caltech, JPL, NHSC,
Univ. Colorado (USA). This development has been supported by national
funding agencies: CSA (Canada); NAOC (China); CEA, CNES, CNRS
(France); ASI (Italy); MCINN (Spain); SNSB (Sweden); STFC, UKSA (UK);
and NASA (USA).  This research used the facilities of the Canadian
Astronomy Data Centre operated by the National Research Council of
Canada with the support of the Canadian Space Agency.

\bibliographystyle{apj}
\bibliography{debrisdisks}

\begin{thebibliography}{}
\expandafter\ifx\csname natexlab\endcsname\relax\def\natexlab#1{#1}\fi

\bibitem[{{Allard} {et~al.}(2012){Allard}, {Homeier}, {Freytag}, \&
  {Sharp}}]{allard2012}
{Allard}, F., {Homeier}, D., {Freytag}, B., \& {Sharp}, C.~M. 2012, in EAS
  Publications Series, Vol.~57, EAS Publications Series, ed. C.~{Reyl{\'e}},
  C.~{Charbonnel}, \& M.~{Schultheis}, 3--43

\bibitem[{{Anglada-Escud{\'e}} \& {Tuomi}(2012)}]{anglada12}
{Anglada-Escud{\'e}}, G., \& {Tuomi}, M. 2012, \aap, 548, A58

\bibitem[{Augereau \& Beust(2006)}]{augereau-beust-2006}
Augereau, J., \& Beust, H. 2006, \aap, 455, 987

\bibitem[{{Augereau} {et~al.}(1999){Augereau}, {Lagrange}, {Mouillet},
  {Papaloizou}, \& {Grorod}}]{augereau-etal-1999}
{Augereau}, J.~C., {Lagrange}, A.~M., {Mouillet}, D., {Papaloizou}, J.~C.~B.,
  \& {Grorod}, P.~A. 1999, \aap, 348, 557

\bibitem[{{Augereau} {et~al.}(2001){Augereau}, {Nelson}, {Lagrange},
  {Papaloizou}, \& {Mouillet}}]{augereau01}
{Augereau}, J.-C., {Nelson}, R.~P., {Lagrange}, A.~M., {Papaloizou}, J.~C.~B.,
  \& {Mouillet}, D. 2001, \aap, 370, 447

\bibitem[{{Balog} {et~al.}(2014){Balog}, {M{\"u}ller}, {Nielbock}, {Altieri},
  {Klaas}, {Blommaert}, {Linz}, {Lutz}, {Mo{\'o}r}, {Billot}, {Sauvage}, \&
  {Okumura}}]{balog14}
{Balog}, Z., {M{\"u}ller}, T., {Nielbock}, M., {et~al.} 2014, Experimental
  Astronomy, 37, 129

\bibitem[{{Binks} \& {Jeffries}(2014)}]{binks14}
{Binks}, A.~S., \& {Jeffries}, R.~D. 2014, \mnras, 438, L11

\bibitem[{{Booth} {et~al.}(2013){Booth}, {Kennedy}, {Sibthorpe}, {Matthews},
  {Wyatt}, {Duch{\^e}ne}, {Kavelaars}, {Rodriguez}, {Greaves}, {Koning},
  {Vican}, {Rieke}, {Su}, {Moro-Mart{\'{\i}}n}, \& {Kalas}}]{booth13}
{Booth}, M., {Kennedy}, G., {Sibthorpe}, B., {et~al.} 2013, \mnras, 428, 1263

\bibitem[{{Burns} {et~al.}(1979){Burns}, {Lamy}, \& {Soter}}]{burns79}
{Burns}, J.~A., {Lamy}, P.~L., \& {Soter}, S. 1979, Icarus, 40, 1

\bibitem[{{Cavanagh} {et~al.}(2008){Cavanagh}, {Jenness}, {Economou}, \&
  {Currie}}]{cavanagh08}
{Cavanagh}, B., {Jenness}, T., {Economou}, F., \& {Currie}, M.~J. 2008,
  Astronomische Nachrichten, 329, 295

\bibitem[{{Chambers}(2014)}]{chambers14}
{Chambers}, J.~E. 2014, Icarus, 233, 83

\bibitem[{{Chapin} {et~al.}(2013){Chapin}, {Berry}, {Gibb}, {Jenness}, {Scott},
  {Tilanus}, {Economou}, \& {Holland}}]{chapin13}
{Chapin}, E.~L., {Berry}, D.~S., {Gibb}, A.~G., {et~al.} 2013, \mnras, 430,
  2545

\bibitem[{{Chen} {et~al.}(2011){Chen}, {Mamajek}, {Bitner}, {Pecaut}, {Su}, \&
  {Weinberger}}]{chen11}
{Chen}, C.~H., {Mamajek}, E.~E., {Bitner}, M.~A., {et~al.} 2011, \apj, 738, 122

\bibitem[{{Chen} {et~al.}(2012){Chen}, {Pecaut}, {Mamajek}, {Su}, \&
  {Bitner}}]{chen12}
{Chen}, C.~H., {Pecaut}, M., {Mamajek}, E.~E., {Su}, K.~Y.~L., \& {Bitner}, M.
  2012, \apj, 756, 133

\bibitem[{{Chen} {et~al.}(2005){Chen}, {Patten}, {Werner}, {Dowell},
  {Stapelfeldt}, {Song}, {Stauffer}, {Blaylock}, {Gordon}, \&
  {Krause}}]{chen05}
{Chen}, C.~H., {Patten}, B.~M., {Werner}, M.~W., {et~al.} 2005, \apj, 634, 1372

\bibitem[{{Cranmer} {et~al.}(2013){Cranmer}, {Wilner}, \&
  {MacGregor}}]{cranmer13}
{Cranmer}, S.~R., {Wilner}, D.~J., \& {MacGregor}, M.~A. 2013, \apj, 772, 149

\bibitem[{{Cully} {et~al.}(1994){Cully}, {Fisher}, {Abbott}, \&
  {Siegmund}}]{cully94}
{Cully}, S.~L., {Fisher}, G.~H., {Abbott}, M.~J., \& {Siegmund}, O.~H.~W. 1994,
  \apj, 435, 449

\bibitem[{{Dempsey} {et~al.}(2013){Dempsey}, {Friberg}, {Jenness}, {Tilanus},
  {Thomas}, {Holland}, {Bintley}, {Berry}, {Chapin}, {Chrysostomou}, {Davis},
  {Gibb}, {Parsons}, \& {Robson}}]{dempsey13}
{Dempsey}, J.~T., {Friberg}, P., {Jenness}, T., {et~al.} 2013, \mnras, 430,
  2534

\bibitem[{{Dent} {et~al.}(2014){Dent}, {Wyatt}, {Roberge}, {Augereau},
  {Casassus}, {Corder}, {Greaves}, {de Gregorio-Monsalvo}, {Hales}, {Jackson},
  {Hughes}, {Lagrange}, {Matthews}, \& {Wilner}}]{dent13}
{Dent}, W.~R.~F., {Wyatt}, M.~C., {Roberge}, A., {et~al.} 2014, Science, 343,
  1490

\bibitem[{{Eiroa} {et~al.}(2013){Eiroa}, {Marshall}, {Mora}, {Montesinos},
  {Absil}, {Augereau}, {Bayo}, {Bryden}, {Danchi}, {del Burgo}, {Ertel},
  {Fridlund}, {Heras}, {Krivov}, {Launhardt}, {Liseau}, {L{\"o}hne},
  {Maldonado}, {Pilbratt}, {Roberge}, {Rodmann}, {Sanz-Forcada}, {Solano},
  {Stapelfeldt}, {Th{\'e}bault}, {Wolf}, {Ardila}, {Ar{\'e}valo}, {Beichmann},
  {Faramaz}, {Gonz{\'a}lez-Garc{\'{\i}}a}, {Guti{\'e}rrez}, {Lebreton},
  {Mart{\'{\i}}nez-Arn{\'a}iz}, {Meeus}, {Montes}, {Olofsson}, {Su}, {White},
  {Barrado}, {Fukagawa}, {Gr{\"u}n}, {Kamp}, {Lorente}, {Morbidelli},
  {M{\"u}ller}, {Mutschke}, {Nakagawa}, {Ribas}, \&
  {Walker}}]{eiroa-et-al-2013}
{Eiroa}, C., {Marshall}, J.~P., {Mora}, A., {et~al.} 2013, \aap, 555, A11

\bibitem[{{Fitzgerald} {et~al.}(2007){Fitzgerald}, {Kalas}, {Duch{\^e}ne},
  {Pinte}, \& {Graham}}]{fitzgerald07}
{Fitzgerald}, M.~P., {Kalas}, P.~G., {Duch{\^e}ne}, G., {Pinte}, C., \&
  {Graham}, J.~R. 2007, \apj, 670, 536

\bibitem[{{Forbrich} {et~al.}(2008){Forbrich}, {Lada}, {Muench}, \&
  {Teixeira}}]{forbrich08}
{Forbrich}, J., {Lada}, C.~J., {Muench}, A.~A., \& {Teixeira}, P.~S. 2008,
  \apj, 687, 1107

\bibitem[{{G{\'a}sp{\'a}r} {et~al.}(2012){G{\'a}sp{\'a}r}, {Psaltis}, {Rieke},
  \& {{\"O}zel}}]{gaspar12}
{G{\'a}sp{\'a}r}, A., {Psaltis}, D., {Rieke}, G.~H., \& {{\"O}zel}, F. 2012,
  \apj, 754, 74

\bibitem[{{Gautier} {et~al.}(2007){Gautier}, {Rieke}, {Stansberry}, {Bryden},
  {Stapelfeldt}, {Werner}, {Beichman}, {Chen}, {Su}, {Trilling}, {Patten}, \&
  {Roellig}}]{gautier07}
{Gautier}, III, T.~N., {Rieke}, G.~H., {Stansberry}, J., {et~al.} 2007, \apj,
  667, 527

\bibitem[{{Gorlova} {et~al.}(2006){Gorlova}, {Rieke}, {Muzerolle}, {Stauffer},
  {Siegler}, {Young}, \& {Stansberry}}]{gorlova06}
{Gorlova}, N., {Rieke}, G.~H., {Muzerolle}, J., {et~al.} 2006, \apj, 649, 1028

\bibitem[{{Graham} {et~al.}(2007){Graham}, {Kalas}, \& {Matthews}}]{graham07}
{Graham}, J.~R., {Kalas}, P.~G., \& {Matthews}, B.~C. 2007, \apj, 654, 595

\bibitem[{{Greaves} {et~al.}(2012){Greaves}, {Hales}, {Mason}, \&
  {Matthews}}]{greaves12}
{Greaves}, J.~S., {Hales}, A.~S., {Mason}, B.~S., \& {Matthews}, B.~C. 2012,
  \mnras, 423, L70

\bibitem[{{Griffin} {et~al.}(2010){Griffin}, {Abergel}, {Abreu}, {Ade},
  {Andr{\'e}}, {Augueres}, {Babbedge}, {Bae}, {Baillie}, {Baluteau}, {Barlow},
  {Bendo}, {Benielli}, {Bock}, {Bonhomme}, {Brisbin}, {Brockley-Blatt},
  {Caldwell}, {Cara}, {Castro-Rodriguez}, {Cerulli}, {Chanial}, {Chen},
  {Clark}, {Clements}, {Clerc}, {Coker}, {Communal}, {Conversi}, {Cox},
  {Crumb}, {Cunningham}, {Daly}, {Davis}, {de Antoni}, {Delderfield}, {Devin},
  {di Giorgio}, {Didschuns}, {Dohlen}, {Donati}, {Dowell}, {Dowell}, {Duband},
  {Dumaye}, {Emery}, {Ferlet}, {Ferrand}, {Fontignie}, {Fox}, {Franceschini},
  {Frerking}, {Fulton}, {Garcia}, {Gastaud}, {Gear}, {Glenn}, {Goizel},
  {Griffin}, {Grundy}, {Guest}, {Guillemet}, {Hargrave}, {Harwit}, {Hastings},
  {Hatziminaoglou}, {Herman}, {Hinde}, {Hristov}, {Huang}, {Imhof}, {Isaak},
  {Israelsson}, {Ivison}, {Jennings}, {Kiernan}, {King}, {Lange}, {Latter},
  {Laurent}, {Laurent}, {Leeks}, {Lellouch}, {Levenson}, {Li}, {Li},
  {Lilienthal}, {Lim}, {Liu}, {Lu}, {Madden}, {Mainetti}, {Marliani}, {McKay},
  {Mercier}, {Molinari}, {Morris}, {Moseley}, {Mulder}, {Mur}, {Naylor},
  {Nguyen}, {O'Halloran}, {Oliver}, {Olofsson}, {Olofsson}, {Orfei}, {Page},
  {Pain}, {Panuzzo}, {Papageorgiou}, {Parks}, {Parr-Burman}, {Pearce},
  {Pearson}, {P{\'e}rez-Fournon}, {Pinsard}, {Pisano}, {Podosek}, {Pohlen},
  {Polehampton}, {Pouliquen}, {Rigopoulou}, {Rizzo}, {Roseboom}, {Roussel},
  {Rowan-Robinson}, {Rownd}, {Saraceno}, {Sauvage}, {Savage}, {Savini},
  {Sawyer}, {Scharmberg}, {Schmitt}, {Schneider}, {Schulz}, {Schwartz},
  {Shafer}, {Shupe}, {Sibthorpe}, {Sidher}, {Smith}, {Smith}, {Smith},
  {Spencer}, {Stobie}, {Sudiwala}, {Sukhatme}, {Surace}, {Stevens}, {Swinyard},
  {Trichas}, {Tourette}, {Triou}, {Tseng}, {Tucker}, {Turner}, {Vaccari},
  {Valtchanov}, {Vigroux}, {Virique}, {Voellmer}, {Walker}, {Ward}, {Waskett},
  {Weilert}, {Wesson}, {White}, {Whitehouse}, {Wilson}, {Winter}, {Woodcraft},
  {Wright}, {Xu}, {Zavagno}, {Zemcov}, {Zhang}, \& {Zonca}}]{griffin10}
{Griffin}, M.~J., {Abergel}, A., {Abreu}, A., {et~al.} 2010, \aap, 518, L3

\bibitem[{{Holland} {et~al.}(1998){Holland}, {Greaves}, {Zuckerman}, {Webb},
  {McCarthy}, {Coulson}, {Walther}, {Dent}, {Gear}, \& {Robson}}]{holland98}
{Holland}, W.~S., {Greaves}, J.~S., {Zuckerman}, B., {et~al.} 1998, \nat, 392,
  788

\bibitem[{{Holland} {et~al.}(2013){Holland}, {Bintley}, {Chapin},
  {Chrysostomou}, {Davis}, {Dempsey}, {Duncan}, {Fich}, {Friberg}, {Halpern},
  {Irwin}, {Jenness}, {Kelly}, {MacIntosh}, {Robson}, {Scott}, {Ade},
  {Atad-Ettedgui}, {Berry}, {Craig}, {Gao}, {Gibb}, {Hilton}, {Hollister},
  {Kycia}, {Lunney}, {McGregor}, {Montgomery}, {Parkes}, {Tilanus}, {Ullom},
  {Walther}, {Walton}, {Woodcraft}, {Amiri}, {Atkinson}, {Burger}, {Chuter},
  {Coulson}, {Doriese}, {Dunare}, {Economou}, {Niemack}, {Parsons},
  {Reintsema}, {Sibthorpe}, {Smail}, {Sudiwala}, \& {Thomas}}]{holland13}
{Holland}, W.~S., {Bintley}, D., {Chapin}, E.~L., {et~al.} 2013, \mnras, 430,
  2513

\bibitem[{{Kalas} {et~al.}(2004){Kalas}, {Liu}, \& {Matthews}}]{kalas04}
{Kalas}, P., {Liu}, M.~C., \& {Matthews}, B.~C. 2004, Science, 303, 1990

\bibitem[{{Kennedy} \& {Wyatt}(2010)}]{kennedy-wyatt-2010}
{Kennedy}, G.~M., \& {Wyatt}, M.~C. 2010, \mnras, 405, 1253

\bibitem[{{Kennedy} {et~al.}(2014){Kennedy}, {Wyatt}, {Kalas}, {Duch{\^e}ne},
  {Sibthorpe}, {Lestrade}, {Matthews}, \& {Greaves}}]{kennedy2013fc}
{Kennedy}, G.~M., {Wyatt}, M.~C., {Kalas}, P., {et~al.} 2014, \mnras, 438, L96

\bibitem[{{Kennedy} {et~al.}(2012{\natexlab{a}}){Kennedy}, {Wyatt},
  {Sibthorpe}, {Phillips}, {Matthews}, \& {Greaves}}]{kennedy12binary}
{Kennedy}, G.~M., {Wyatt}, M.~C., {Sibthorpe}, B., {et~al.} 2012{\natexlab{a}},
  \mnras, 426, 2115

\bibitem[{{Kennedy} {et~al.}(2012{\natexlab{b}}){Kennedy}, {Wyatt},
  {Sibthorpe}, {Duch{\^e}ne}, {Kalas}, {Matthews}, {Greaves}, {Su}, \&
  {Fitzgerald}}]{kennedy12_99her}
---. 2012{\natexlab{b}}, \mnras, 421, 2264

\bibitem[{{Kenyon} \& {Bromley}(2002)}]{kenyon-bromley-2002}
{Kenyon}, S.~J., \& {Bromley}, B.~C. 2002, \aj, 123, 1757

\bibitem[{{Krist} {et~al.}(2005){Krist}, {Ardila}, {Golimowski}, {Clampin},
  {Ford}, {Illingworth}, {Hartig}, {Bartko}, {Ben{\'{\i}}tez}, {Blakeslee},
  {Bouwens}, {Bradley}, {Broadhurst}, {Brown}, {Burrows}, {Cheng}, {Cross},
  {Demarco}, {Feldman}, {Franx}, {Goto}, {Gronwall}, {Holden}, {Homeier},
  {Infante}, {Kimble}, {Lesser}, {Martel}, {Mei}, {Menanteau}, {Meurer},
  {Miley}, {Motta}, {Postman}, {Rosati}, {Sirianni}, {Sparks}, {Tran},
  {Tsvetanov}, {White}, \& {Zheng}}]{krist05}
{Krist}, J.~E., {Ardila}, D.~R., {Golimowski}, D.~A., {et~al.} 2005, \aj, 129,
  1008

\bibitem[{{Lestrade} {et~al.}(2006){Lestrade}, {Wyatt}, {Bertoldi}, {Dent}, \&
  {Menten}}]{lestrade06}
{Lestrade}, J.-F., {Wyatt}, M.~C., {Bertoldi}, F., {Dent}, W.~R.~F., \&
  {Menten}, K.~M. 2006, \aap, 460, 733

\bibitem[{{Lestrade} {et~al.}(2012){Lestrade}, {Matthews}, {Sibthorpe},
  {Kennedy}, {Wyatt}, {Bryden}, {Greaves}, {Thilliez}, {Moro-Mart{\'{\i}}n},
  {Booth}, {Dent}, {Duch{\^e}ne}, {Harvey}, {Horner}, {Kalas}, {Kavelaars},
  {Phillips}, {Rodriguez}, {Su}, \& {Wilner}}]{lestrade12}
{Lestrade}, J.-F., {Matthews}, B.~C., {Sibthorpe}, B., {et~al.} 2012, \aap,
  548, A86

\bibitem[{{Li} \& {Greenberg}(1997)}]{li-greenberg-1997}
{Li}, A., \& {Greenberg}, J.~M. 1997, \aap, 323, 566

\bibitem[{{Lim} \& {White}(1996)}]{lim96}
{Lim}, J., \& {White}, S.~M. 1996, \apjl, 462, L91

\bibitem[{{Liu}(2004)}]{liu04alone}
{Liu}, M.~C. 2004, Science, 305, 1442

\bibitem[{{Liu} {et~al.}(2004){Liu}, {Matthews}, {Williams}, \&
  {Kalas}}]{liu04}
{Liu}, M.~C., {Matthews}, B.~C., {Williams}, J.~P., \& {Kalas}, P.~G. 2004,
  \apj, 608, 526

\bibitem[{{Low} {et~al.}(2005){Low}, {Smith}, {Werner}, {Chen}, {Krause},
  {Jura}, \& {Hines}}]{low05}
{Low}, F.~J., {Smith}, P.~S., {Werner}, M., {et~al.} 2005, \apj, 631, 1170

\bibitem[{{MacGregor} {et~al.}(2013){MacGregor}, {Wilner}, {Rosenfeld},
  {Andrews}, {Matthews}, {Hughes}, {Booth}, {Chiang}, {Graham}, {Kalas},
  {Kennedy}, \& {Sibthorpe}}]{macgregor13}
{MacGregor}, M.~A., {Wilner}, D.~J., {Rosenfeld}, K.~A., {et~al.} 2013, \apjl,
  762, L21

\bibitem[{{Malo} {et~al.}(2014){Malo}, {Doyon}, {Feiden}, {Albert},
  {Lafreni{\`e}re}, {Artigau}, {Gagn{\'e}}, \& {Riedel}}]{malo14}
{Malo}, L., {Doyon}, R., {Feiden}, G.~A., {et~al.} 2014, \apj, 792, 37

\bibitem[{{Mamajek} \& {Bell}(2014)}]{mamajek14}
{Mamajek}, E.~E., \& {Bell}, C.~P.~M. 2014, \mnras, 445, 2169

\bibitem[{{Mamajek} {et~al.}(2013){Mamajek}, {Bartlett}, {Seifahrt}, {Henry},
  {Dieterich}, {Lurie}, {Kenworthy}, {Jao}, {Riedel}, {Subasavage}, {Winters},
  {Finch}, {Ianna}, \& {Bean}}]{mamajek13}
{Mamajek}, E.~E., {Bartlett}, J.~L., {Seifahrt}, A., {et~al.} 2013, \aj, 146,
  154

\bibitem[{{Matthews} {et~al.}(2014{\natexlab{a}}){Matthews}, {Kennedy},
  {Sibthorpe}, {Booth}, {Wyatt}, {Broekhoven-Fiene}, {Macintosh}, \&
  {Marois}}]{matthews14}
{Matthews}, B., {Kennedy}, G., {Sibthorpe}, B., {et~al.} 2014{\natexlab{a}},
  \apj, 780, 97

\bibitem[{{Matthews} {et~al.}(2007){Matthews}, {Kalas}, \&
  {Wyatt}}]{matthews07}
{Matthews}, B.~C., {Kalas}, P.~G., \& {Wyatt}, M.~C. 2007, \apj, 663, 1103

\bibitem[{{Matthews} {et~al.}(2014{\natexlab{b}}){Matthews}, {Krivov}, {Wyatt},
  {Bryden}, \& {Eiroa}}]{matthews14_ppvi}
{Matthews}, B.~C., {Krivov}, A.~V., {Wyatt}, M.~C., {Bryden}, G., \& {Eiroa},
  C. 2014{\natexlab{b}}, Protostars and Planets VI, 521

\bibitem[{{Matthews} {et~al.}(2015){Matthews}, {Sibthorpe}, {Kennedy}, {Wyatt},
  {Booth}, {Greaves}, \& {Phillips}}]{matthews15}
{Matthews}, B.~C., {Sibthorpe}, B., {Kennedy}, G., {et~al.} 2015, \mnras, in
  preparation

\bibitem[{{Metchev} {et~al.}(2005){Metchev}, {Eisner}, {Hillenbrand}, \&
  {Wolf}}]{metchev-etal-2005}
{Metchev}, S.~A., {Eisner}, J.~A., {Hillenbrand}, L.~A., \& {Wolf}, S. 2005,
  \apj, 622, 451

\bibitem[{{Moshir} \& {et al.}(1990)}]{moshir1990}
{Moshir}, M., \& {et al.} 1990, in IRAS Faint Source Catalogue, version 2.0
  (1990), 0

\bibitem[{{Ott}(2010)}]{ott10}
{Ott}, S. 2010, in Astronomical Society of the Pacific Conference Series, Vol.
  434, Astronomical Data Analysis Software and Systems XIX, ed. Y.~{Mizumoto},
  K.-I. {Morita}, \& M.~{Ohishi}, 139

\bibitem[{{Pantin} {et~al.}(1997){Pantin}, {Lagage}, \&
  {Artymowicz}}]{pantin97}
{Pantin}, E., {Lagage}, P.~O., \& {Artymowicz}, P. 1997, \aap, 327, 1123

\bibitem[{{Pecaut} \& {Mamajek}(2013)}]{pecaut-mamajek-2013}
{Pecaut}, M.~J., \& {Mamajek}, E.~E. 2013, \apjs, 208, 9

\bibitem[{{Phillips} {et~al.}(2010){Phillips}, {Greaves}, {Dent}, {Matthews},
  {Holland}, {Wyatt}, \& {Sibthorpe}}]{phillips10}
{Phillips}, N.~M., {Greaves}, J.~S., {Dent}, W.~R.~F., {et~al.} 2010, \mnras,
  403, 1089

\bibitem[{{Pilbratt} {et~al.}(2010){Pilbratt}, {Riedinger}, {Passvogel},
  {Crone}, {Doyle}, {Gageur}, {Heras}, {Jewell}, {Metcalfe}, {Ott}, \&
  {Schmidt}}]{pilbratt10}
{Pilbratt}, G.~L., {Riedinger}, J.~R., {Passvogel}, T., {et~al.} 2010, \aap,
  518, L1

\bibitem[{{Plavchan} {et~al.}(2005){Plavchan}, {Jura}, \&
  {Lipscy}}]{plavchan-et-al-2005}
{Plavchan}, P., {Jura}, M., \& {Lipscy}, S.~J. 2005, \apj, 631, 1161

\bibitem[{{Plavchan} {et~al.}(2009){Plavchan}, {Werner}, {Chen}, {Stapelfeldt},
  {Su}, {Stauffer}, \& {Song}}]{plavchan09}
{Plavchan}, P., {Werner}, M.~W., {Chen}, C.~H., {et~al.} 2009, \apj, 698, 1068

\bibitem[{{Poglitsch} {et~al.}(2010){Poglitsch}, {Waelkens}, {Geis},
  {Feuchtgruber}, {Vandenbussche}, {Rodriguez}, {Krause}, {Renotte}, {van
  Hoof}, {Saraceno}, {Cepa}, {Kerschbaum}, {Agn{\`e}se}, {Ali}, {Altieri},
  {Andreani}, {Augueres}, {Balog}, {Barl}, {Bauer}, {Belbachir}, {Benedettini},
  {Billot}, {Boulade}, {Bischof}, {Blommaert}, {Callut}, {Cara}, {Cerulli},
  {Cesarsky}, {Contursi}, {Creten}, {De Meester}, {Doublier}, {Doumayrou},
  {Duband}, {Exter}, {Genzel}, {Gillis}, {Gr{\"o}zinger}, {Henning},
  {Herreros}, {Huygen}, {Inguscio}, {Jakob}, {Jamar}, {Jean}, {de Jong},
  {Katterloher}, {Kiss}, {Klaas}, {Lemke}, {Lutz}, {Madden}, {Marquet},
  {Martignac}, {Mazy}, {Merken}, {Montfort}, {Morbidelli}, {M{\"u}ller},
  {Nielbock}, {Okumura}, {Orfei}, {Ottensamer}, {Pezzuto}, {Popesso},
  {Putzeys}, {Regibo}, {Reveret}, {Royer}, {Sauvage}, {Schreiber}, {Stegmaier},
  {Schmitt}, {Schubert}, {Sturm}, {Thiel}, {Tofani}, {Vavrek}, {Wetzstein},
  {Wieprecht}, \& {Wiezorrek}}]{poglitsch10}
{Poglitsch}, A., {Waelkens}, C., {Geis}, N., {et~al.} 2010, \aap, 518, L2

\bibitem[{{Raymond} {et~al.}(2012){Raymond}, {Armitage}, {Moro-Mart{\'{\i}}n},
  {Booth}, {Wyatt}, {Armstrong}, {Mandell}, {Selsis}, \& {West}}]{raymond12}
{Raymond}, S.~N., {Armitage}, P.~J., {Moro-Mart{\'{\i}}n}, A., {et~al.} 2012,
  \aap, 541, A11

\bibitem[{{Rebull} {et~al.}(2008){Rebull}, {Stapelfeldt}, {Werner}, {Mannings},
  {Chen}, {Stauffer}, {Smith}, {Song}, {Hines}, \& {Low}}]{rebull08}
{Rebull}, L.~M., {Stapelfeldt}, K.~R., {Werner}, M.~W., {et~al.} 2008, \apj,
  681, 1484

\bibitem[{{Rivera} {et~al.}(2005){Rivera}, {Lissauer}, {Butler}, {Marcy},
  {Vogt}, {Fischer}, {Brown}, {Laughlin}, \& {Henry}}]{rivera05}
{Rivera}, E.~J., {Lissauer}, J.~J., {Butler}, R.~P., {et~al.} 2005, \apj, 634,
  625

\bibitem[{{Schneider} {et~al.}(2014){Schneider}, {Grady}, {Hines}, {Stark},
  {Debes}, {Carson}, {Kuchner}, {Perrin}, {Weinberger}, {Wisniewski},
  {Silverstone}, {Jang-Condell}, {Henning}, {Woodgate}, {Serabyn},
  {Moro-Martin}, {Tamura}, {Hinz}, \& {Rodigas}}]{schneider14}
{Schneider}, G., {Grady}, C.~A., {Hines}, D.~C., {et~al.} 2014, \aj, 148, 59

\bibitem[{{Sch{\"u}ppler} {et~al.}(2015){Sch{\"u}ppler}, {L{\"o}hne}, {Krivov},
  {Ertel}, {Marshall}, {Wolf}, {Wyatt}, {Augereau}, \&
  {Metchev}}]{schueppler2015}
{Sch{\"u}ppler}, C., {L{\"o}hne}, T., {Krivov}, A.~V., {et~al.} 2015, ArXiv
  e-prints, arXiv:1506.04564

\bibitem[{{Shen} {et~al.}(2009){Shen}, {Draine}, \& {Johnson}}]{shen09}
{Shen}, Y., {Draine}, B.~T., \& {Johnson}, E.~T. 2009, \apj, 696, 2126

\bibitem[{{Sibthorpe} {et~al.}(2010){Sibthorpe}, {Vandenbussche}, {Greaves},
  {Pantin}, {Olofsson}, {Acke}, {Barlow}, {Blommaert}, {Bouwman}, {Brandeker},
  {Cohen}, {De Meester}, {Dent}, {di Francesco}, {Dominik}, {Fridlund}, {Gear},
  {Glauser}, {Gomez}, {Hargrave}, {Harvey}, {Henning}, {Heras}, {Hogerheijde},
  {Holland}, {Ivison}, {Leeks}, {Lim}, {Liseau}, {Matthews}, {Naylor},
  {Pilbratt}, {Polehampton}, {Regibo}, {Royer}, {Sicilia-Aguilar}, {Swinyard},
  {Waelkens}, {Walker}, \& {Wesson}}]{sibthorpe10}
{Sibthorpe}, B., {Vandenbussche}, B., {Greaves}, J.~S., {et~al.} 2010, \aap,
  518, L130

\bibitem[{{Sierchio} {et~al.}(2010){Sierchio}, {Rieke}, {Su}, {Plavchan},
  {Stauffer}, \& {Gorlova}}]{sierchio10}
{Sierchio}, J.~M., {Rieke}, G.~H., {Su}, K.~Y.~L., {et~al.} 2010, \apj, 712,
  1421

\bibitem[{Strubbe \& Chiang(2006)}]{strubbe-chiang-2006}
Strubbe, L.~E., \& Chiang, E.~I. 2006, \apj, 648, 652

\bibitem[{{Su} {et~al.}(2006){Su}, {Rieke}, {Stansberry}, {Bryden},
  {Stapelfeldt}, {Trilling}, {Muzerolle}, {Beichman}, {Moro-Martin}, {Hines},
  \& {Werner}}]{su06}
{Su}, K.~Y.~L., {Rieke}, G.~H., {Stansberry}, J.~A., {et~al.} 2006, \apj, 653,
  675

\bibitem[{{Su} {et~al.}(2009){Su}, {Rieke}, {Stapelfeldt}, {Malhotra},
  {Bryden}, {Smith}, {Misselt}, {Moro-Martin}, \& {Williams}}]{su09}
{Su}, K.~Y.~L., {Rieke}, G.~H., {Stapelfeldt}, K.~R., {et~al.} 2009, \apj, 705,
  314

\bibitem[{{Thureau} {et~al.}(2014){Thureau}, {Greaves}, {Matthews}, {Kennedy},
  {Phillips}, {Booth}, {Duch{\^e}ne}, {Horner}, {Rodriguez}, {Sibthorpe}, \&
  {Wyatt}}]{thureau13}
{Thureau}, N.~D., {Greaves}, J.~S., {Matthews}, B.~C., {et~al.} 2014, \mnras,
  445, 2558

\bibitem[{{Udry} {et~al.}(2007){Udry}, {Bonfils}, {Delfosse}, {Forveille},
  {Mayor}, {Perrier}, {Bouchy}, {Lovis}, {Pepe}, {Queloz}, \&
  {Bertaux}}]{udry07}
{Udry}, S., {Bonfils}, X., {Delfosse}, X., {et~al.} 2007, \aap, 469, L43

\bibitem[{{van Leeuwen}(2007)}]{vanleeuwen07}
{van Leeuwen}, F., ed. 2007, Astrophysics and Space Science Library, Vol. 350,
  {Hipparcos, the New Reduction of the Raw Data}

\bibitem[{{Williams} \& {Andrews}(2006)}]{williams-andrews-2006}
{Williams}, J.~P., \& {Andrews}, S.~M. 2006, \apj, 653, 1480

\bibitem[{{Wilner} {et~al.}(2012){Wilner}, {Andrews}, {MacGregor}, \&
  {Hughes}}]{wilner12}
{Wilner}, D.~J., {Andrews}, S.~M., {MacGregor}, M.~A., \& {Hughes}, A.~M. 2012,
  \apjl, 749, L27

\bibitem[{{Wyatt}(2008)}]{wyatt-2008}
{Wyatt}, M.~C. 2008, ARAA, 46, 339

\bibitem[{{Wyatt} \& {Dent}(2002)}]{Wyatt02}
{Wyatt}, M.~C., \& {Dent}, W.~R.~F. 2002, \mnras, 334, 589

\bibitem[{{Zuckerman} {et~al.}(2001){Zuckerman}, {Song}, {Bessell}, \&
  {Webb}}]{zuckerman01}
{Zuckerman}, B., {Song}, I., {Bessell}, M.~S., \& {Webb}, R.~A. 2001, \apjl,
  562, L87

\end{thebibliography}

\end{document}